%% file: main.tex
\documentclass[a4paper,fleqn,usenatbib,useAMS]{mnras}

\usepackage{graphicx}	
\usepackage{amsmath}	
\usepackage{amssymb}	
\usepackage{multicol}   
\usepackage{bm}	  	    

\usepackage{xspace}
\newcommand{\cmc}{\,cm$^{-3}$\xspace}
\newcommand{\muG}{\,$\mu$G\xspace}

\newcommand{\kms}{\,km s$^{-1}$\xspace}
\newcommand{\OII}{\,[\ion{O}{ii}]\xspace}
\newcommand{\OIII}{\,[\ion{O}{iii}]\xspace}
\newcommand{\OIV}{\,[\ion{O}{iv}]\xspace}
\newcommand{\NII}{\,[\ion{N}{ii}]\xspace}

\newcommand{\SII}{\,[\ion{S}{ii}]\xspace}

\newcommand{\Ha}{\,H$\alpha$\xspace}
\newcommand{\Hb}{\,H$\beta$\xspace}

\usepackage[T1]{fontenc}
\usepackage{ae,aecompl}

\usepackage{mathptmx}

\title[IC 443]{A multispectral analysis of the  northeastern shell of IC 443}

\author[Alarie \& Drissen]{
Alexandre Alarie,$^{1,2,3}$
Laurent Drissen,$^{2,3}$
\\
$^{1}$ Instituto de Astronom\'{i}a, Universidad National Aut\'{o}noma de M\'{e}xico, Apdo. Postal 70264, 04510 Mexico D.F., Mexico \\
$^{2}$ D\'{e}partement de physique, de g\'{e}nie physique et d'optique, Universit\'{e} Laval, Qu\'{e}bec, QC, G1V 0A6, Canada \\
$^{3}$ Centre de Recherche en Astrophysique du Qu\'{e}bec\\
}

\pubyear{2019}

\begin{document}
\label{firstpage}
\pagerange{\pageref{firstpage}--\pageref{lastpage}}
\maketitle

\begin{abstract}
We have carried out optical observations of the north-eastern part of the supernova remnant IC 443 using the CFHT imaging spectrograph SITELLE. The observations consist of three multispectral cubes covering an 11$^{\prime}$ $\times$11$^{\prime}$ area allowing the investigation of both the spatial and spectral variation of 9 emission lines : \OII $\lambda\lambda$3726+3729, \OIII $\lambda\lambda$4959,5007, \Hb, \Ha, \NII $\lambda\lambda$6548,6583 and \SII $\lambda\lambda$6716,6731. Extinction measurement from the \Ha/\Hb shows significant variation across the observed region with E(B-V) = 0.8-1.1. Electron density measurements using \SII lines indicate densities ranging from 100 up to 2500 \cmc. Models computed with the shock modelling code \textsc{mappings} are presented and compared with the observations. A combination of complete shock model and truncated ones are required in order to explain the observed spectrum. The shock velocities found in IC 443 are between 20 and 150 \kms with 75 \kms being the most prominent velocity. The pre-shock number density varies from 20 to 60 \cmc. A single set of abundances close to solar values combined with varying shock parameters (shock velocity, pre-shock density and shock age) are sufficient to explain to great variation of lines intensities observed in IC 443. Despite the relatively modest spectral resolution of the data (R$\sim 1500$ at \Ha), we clearly separate the red and blue velocity components of the expanding nebula, which show significant morphological differences.
\end{abstract}



\section{Introduction}

IC 443 (G189.1+3.0) is an evolved galactic supernova remnant (SNR) located in the direction of the galactic anti-center at a distance between 1.5 and 2 kpc \citep{Petre:1988,Welsh:2003,Ambrocio-Cruz:2017} and an age estimated at around 3000-30,000 years \citep{Petre:1988, Olbert:2001}. Because of its large scale ($\approx$0.5$^{\circ}$) and clear view without any obstruction in the direction of the remnant, IC 443 is known to be an excellent laboratory to study the complex interaction between a SNR and the interstellar and circumstellar medium. 

IC 443 has been classified as a mixed-morphology remnant: shell-like and bright structure in radio continuum and in the optical \citep{Lee:2008} and centrally filled in X-ray. \citet{Braun:1986} suggested that the remnant has evolved inside a preexisting wind-blown bubble, which likely was formed by the remnant's massive progenitor star. This proposed explanation was supported by \citet{Troja:2006,Troja:2008} based on the analysis of X-ray observations obtained with XMM-Newton. IC 443 is apparently interacting with both atomic and molecular gas \citep{Snell:2005}. In the northeastern part, the SNR shock is encountering mostly an atomic medium, and only the atomic lines expected in the post-shock recombining gas are detected \citep{Fesen:1980}. The southern boundary, on the other hand, shows various molecular lines with broad wings suggesting that the shock is propagating into mostly molecular medium \citep{Burton:1988,Dickman:1992,Dishoeck:1993}. 

In recent years, there has been considerable interest in studying the complex interaction between the IC 443 SNR with its surrounding molecular clouds at radio and infrared wavelengths \citep{Hezareh:2013,Fang:2013,Su:2014} as well as the determination of the abundances of heavy elements in some part of the remnant from X-ray observations \citep{Troja:2008, Bocchino:2009,Swartz:2015}. 

Detections of high and very high energy gamma rays have also been reported in the vicinity of IC 443, which overlap with the object's structure seen in the radio, IR, optical and X-ray bands. First with the MAGIC telescope \citep{Albert:2007}, and then early in the Fermi-LAT mission \citep{Abdo:2010,Ackermann:2013}. Is is assumed that these emissions are associated with locally accelerated cosmic rays.

Despite all these observations made in recent years, optical spectroscopy, which samples very different physical phenomena and chemical el-
ements,  are limited to long-slit spectroscopy   \citep{Parker:1964,Esipov:1972,Fesen:1980}.

This present work was undertaken to improve the spectroscopic data available for the north-eastern part of IC 443 where optical lines are present. Multi-spectral cubes obtained with the imaging Fourier transform spectrometer SITELLE are presented and analyzed alongside with shock models to infer shock properties. The paper is organized as follows: in section \ref{sec:observations} we present the observations and data reduction; in section \ref{sec:analysis} we discuss the shock modelling technique and make direct comparisons with observations. A kinematical analysis of the data is presented in section  \ref{sec:kinematics}. We conclude this paper with remarks in section \ref{sec:remarks}.

\section{Observations and data reduction}
\label{sec:observations}

\subsection{Observations}

The data presented here consist of three multispectral data cubes obtained with the optical imaging Fourier Transform Spectrometer (iFTS), SITELLE, attached to the 3.6-meter Canada-France-Hawaii Telescope \citep{Drissen:2019}. SITELLE delivers multispectral cubes covering a 11$^{\prime}$ $\times$11$^{\prime}$ field-of-view, sampled with $0.32^{\prime\prime}$ pixels, covering specific bandpasses of the visible band using filters . A single region of IC 443, centered at RA 6h17m36.25s and DEC +22$^{\circ}$49$^{\prime}$15$^{\prime\prime}$,   was observed during the nights of 2016 March 4, 6 \& 7 using three filters targeting bright emission lines : \Ha, \Hb, \OII $\lambda$3726+3729, \OIII $\lambda$4959,5007, \NII $\lambda$6548,6583 and \SII $\lambda$6716,6731. SITELLE's spectral resolution R is adjustable to the need of the observer; we selected relatively modest values of R (see Table \ref{tab:observations}) allowing us to clearly separate the emission lines from their neighbors as our primary goal was not to study the kinematics. However, as we show in section \ref{sec:kinematics}, the high signal-to-noise reached in the SN3 filter (around \Ha) allowed us to perform a preliminary kinematical analysis of the target.

\input{table1.tex}

\subsection{Data reduction and line fitting}
\label{sec:reduction}

The data were reduced using SITELLE's dedicated data reduction software \textsc{orbs} \citep{Martin:2015}. Standard reduction procedures were applied (flat, bias, dark) and the flux calibration was performed using a datacube of the spectrophotometric standard GD71 for each filter to define the transmission function at each wavelength. The spectral calibration was done using a helium-neon laser datacube acquired during daytime. A detailed description regarding the reduction process is provided by \citet{Martin:2017}. 

\begin{figure}
\includegraphics{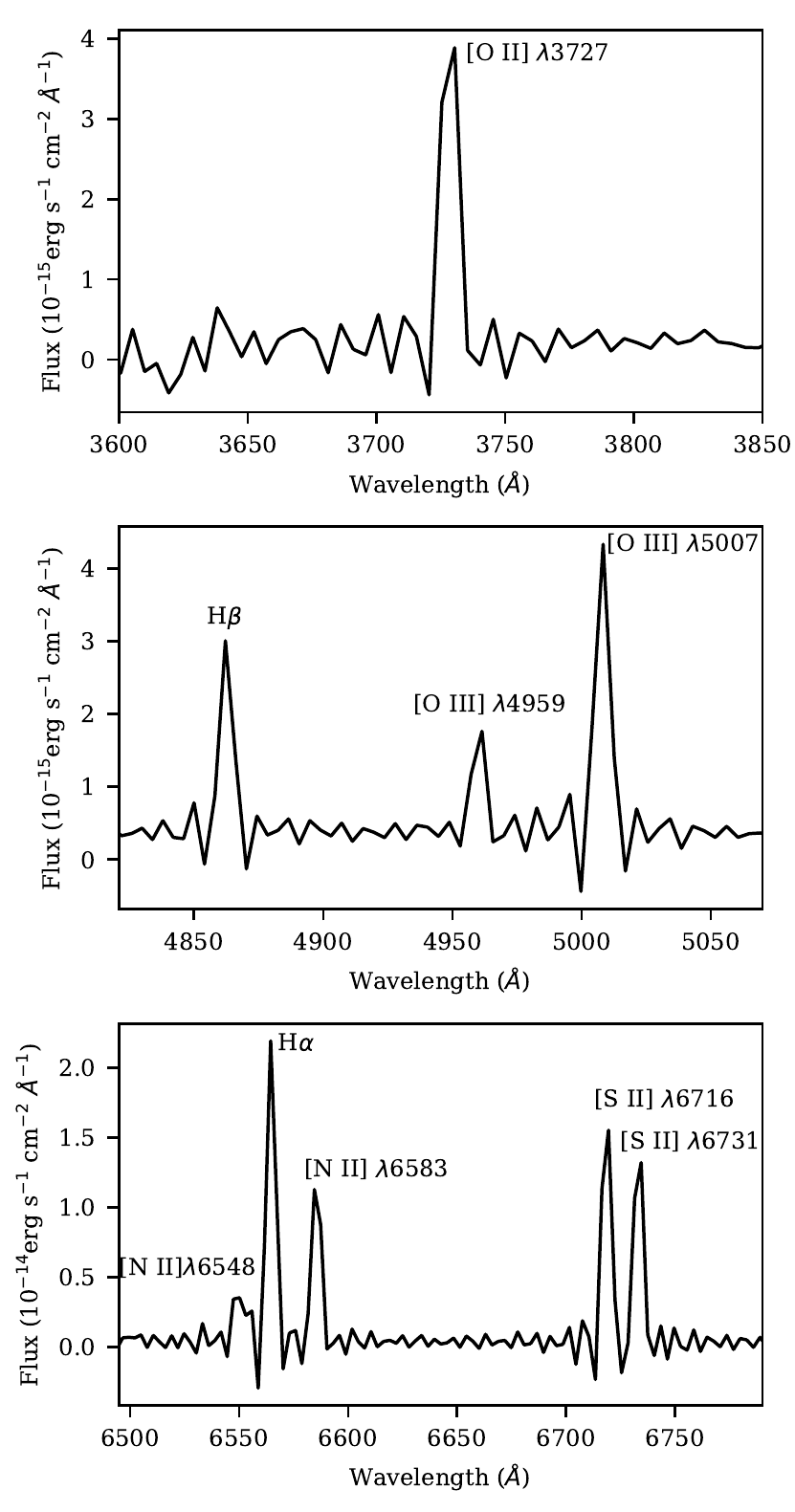}
\caption{Spectra extracted from the data cubes using a circular aperture (2.22$^{\prime\prime}$) centered at RA 6h17m57s and DEC 22$^{\circ}$44$^{\prime}$42$^{\prime\prime}$.}
\label{fig:spectrum}
\end{figure}

An individual spectrum is shown in Figure \ref{fig:spectrum}. It has to be noted that the apparent wiggles on each side of the emission lines are not due to noise but rather to the cardinal sine function (sinc) typical of all Fourier transform spectrometers, as we chose not to apodize the signal. This peculiar instrument line shape is taken into account by the dedicated line-fitting software \textsc{orcs} \citep{Martin:2016} which was specifically created to deal with SITELLE data and therefore uses sinc-shaped line profiles to accurately fit each emission line. In order to improve the signal to noise ratio, the line flux extraction was performed using a 3$\times$3 pixel binning, except for the kinematical analysis, where no binning was done. This manipulation reduced the spatial resolution to approximately 1$^{\prime\prime}$/pixel. 

\begin{figure}
\includegraphics[width=\columnwidth]{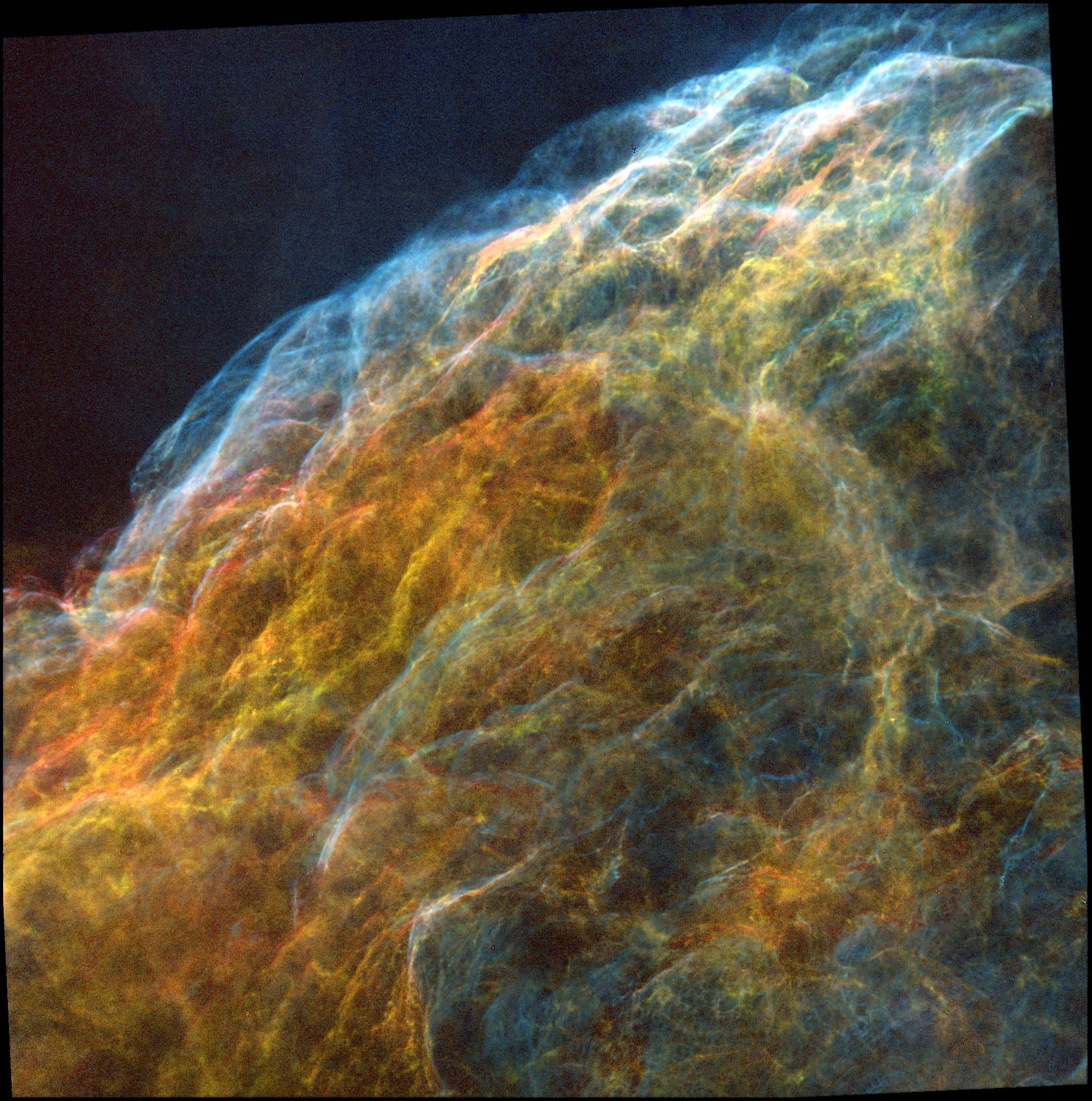}
\caption{Composite image of IC 443, using SITELLE's SN2 and SN3 data cubes. \Ha is coded in red, \SII $\lambda$$\lambda$6716,31
in green and \OIII $\lambda$5007 in blue. The field of view is $11' \times 11'$, with East to the left and North at the top.}
\label{fig:color-image}
\end{figure}

\subsection{Flux and line ratio maps}
\label{sec:line_fitting}

Using the flux and error maps generated by \textsc{orcs}, we have created ratio maps and only kept pixels for which the ratios error is smaller than 10\%. Flux maps in all emission lines are shown the Appendix \ref{sec:figures} : \OII \ref{fig:image OII}, \OIII \ref{fig:image OIII}, \Ha \ref{fig:image Ha}, \NII \ref{fig:image NII} and \SII \ref{fig:image SII}. An unbinned color composite image of the field, using the \Ha, \SII and \OIII flux maps, is shown in Figure \ref{fig:color-image}.

The followings ratios also generated and presented in Appendix \ref{sec:figures} : \OII/\Ha (\ref{fig:colormap_OII_Ha}), \OIII/\Hb (\ref{fig:colormap_OIII_Hb}), \OIII/\OII (\ref{fig:colormap_OIII_OII}), \NII/\Ha (\ref{fig:colormap_NII_Ha}), \SII/\Ha (\ref{fig:colormap_SII_Ha}), \NII/\SII (\ref{fig:colormap_NII_SII}) and \SII $\lambda$6731/$\lambda$6716 (\ref{fig:colormap_SII_SII}). Due to insufficient exposure time during the acquisition of the SN1 cube, \OII emissions could not be accurately measured at many locations leading to incomplete \OII/\Ha and \OIII/\OII maps. 


\subsection{Extinction correction} 

IC 443 is well known to contain patchy obscuration over the whole remnant and its surroundings. Since the observations cover an appreciable fraction of the northeastern shell, it is essential to measure the interstellar reddening at various locations and correct the observed line intensities accordingly. For interstellar radiative shocks, the Balmer lines of hydrogen are thought to be primarily produced by recombination with only a small amount of direct collisional excitation from ground level to bound excited states. Shock models calculated by \citet{Raymond:1979} have shown that the \Ha/\Hb ratio fall between 2.9 and 3.3 with larger values occurring for the slower shocks. Since we do not have the information about the shock velocity prior to a reduction, we used the standard value of \Ha/\Hb of 3.0 to correct our observations. With this value, we estimated the colour excess E(B-V) for every pixel where the error on the \Ha/\Hb ratio is smaller than 10\%. The extinction was evaluated using \textsc{pyneb} \citep{Luridiana:2012} and the \textsc{RedCorr} class which makes use of the \cite{Fitzpatrick:1999} correction curve. Since the extinction is most likely diffuse in front of the remnant, we used a Gaussian blur with a standard deviation of 6$\sigma$ in order to smooth the extinction map and fill regions where the extinction could not be directly measured. The map used to correct the line intensities pixel by pixel is shown in Figure \ref{fig:extinction_map}\footnote{The map can be found in the supplementary material in FITS format.}. The values found are in agreement with earlier measurements, such as those reported by \cite{Fesen:1980}: 
E(B-V) = 0.8 - 1.1.

\subsection{Faint nebulosities from the HII region S249}

Some faint nebulosities visible at the northern edge of our field are due to the H II region S249. These emissions undoubtedly contaminate some of IC 443 spectra overlapping these regions. We estimate that the contamination from the brightest \Ha and \OIII emission from S249 could contribute up to 10\% and 8\% respectively of the measured emission from IC 443. Despite this contamination, we do not attempt the subtract it from the emissions coming from IC 443. Our decision is based on the fact that both the intensity and shape of the emissions coming from S249 behind IC 443 are unpredictable. It has to be noted that the contamination does not seem to affect the whole observed region, but is rather localized in the northern part of IC 443. 

\section{Line ratio analysis}
\label{sec:analysis}

\subsection{Comparison with previous observations}

Very few spectroscopic observations in the optical regime can be found in the literature. The earliest measurements are presented by \cite{Parker:1964} where multiple spectra were taken on the surface of IC 443. Table 2 of their paper shows the line intensities as well as a few line ratios which could be compared with our observations. One position noted as filament 7$^{\prime}$ is located in the north-eastern part of the nebula. The exact location of that filament is unknown since its coordinates were not provided. By approximating its location from their finding charts, we compared visually their \NII/\SII, \SII/\Ha, \NII/\Ha and \OIII/\Hb ratios with our ratio maps and found a very similar set of values for multiple locations around the same area. 

So far, the most extensive spectroscopic analysis in the optical was carried out by \citet{Fesen:1980}. They targetted six positions in IC 443, among which two (their numbers 1 and 4) are located inside our field of view. The positions were sufficiently well indicated to allow a good comparison. As mentioned earlier, their observed \Ha/\Hb ratio is very similar to ours, which lead to the same measurement of the extinction. Also, the line ratios at those places such as \OIII/\Ha, \OII/\Ha, \NII/\Ha, \SII/\Ha and \SII $\lambda$6716/$\lambda$6731 are in very good agreement with our measurements. Furthermore, \citet{Fesen:1980} compared their observations with shock models to investigate both the shock conditions and elemental abundances.

The data presented in this paper can also be compared with shock models in order to investigate shock conditions at the surface of IC 443. Using SITELLE's data with both the spectral and spatial informations, we can attempt to associate spatial line intensity variations directly with shock conditions. The next section is dedicated to the analysis of shock models combined with our observations. Further comparisons are made between the analysis of \citet{Fesen:1980} and previous modelling of shock-wave spectra.  

\subsection{Comparison with new shock models}

Several shock models have been published \citep{Shull:1979,Raymond:1979,Dopita:1996,Allen:2008} and proved useful to study the properties of a great variety of astrophysical objects, especially evolved supernova remnants such as the Cygnus Loop and IC 443. Despite their undeniable usefulness, each grid only covers a very limited set of shock parameters and limit the analysis that can be done for a specific object. To circumvent this problem, we built our own shock models in an attempt to reconcile physical condition found in IC 443 and emission line variations derived from our observations. 

We used the shock modelling code \textsc{mappings v} \citep{Sutherland:2017}. This is an updated version of previous releases: \textsc{mappings iv} \citep{Dopita:2013}, \textsc{mappings iii} \citep{Groves:2004A,Groves:2004B,Allen:2008}, \textsc{mappings ii} \citep{Sutherland:1993} and \textsc{mappings i} \citep{Binette:1982,Dopita:1982,Binette:1985} which were themselves based on an earlier version of the code created by Michael A. Dopita \citep{Dopita:1976,Dopita:1977A,Dopita:1977B,Dopita:1978,Dopita:1979,Dopita:1980}. Details concerning the intricate physics and computational recipes inside the code can be found among the cited papers.

Modelling shock wave spectra require the knowledge of several shock parameters used as input during computation. In the next sections, each parameter is explored independently in order to simplify the overall analysis and discussion. Comparisons with published shock models are made as well as comparisons between the conditions derived from previous analysis on IC 443 and those derived from our observations.

\subsubsection{Preionization}

\begin{figure}
\includegraphics{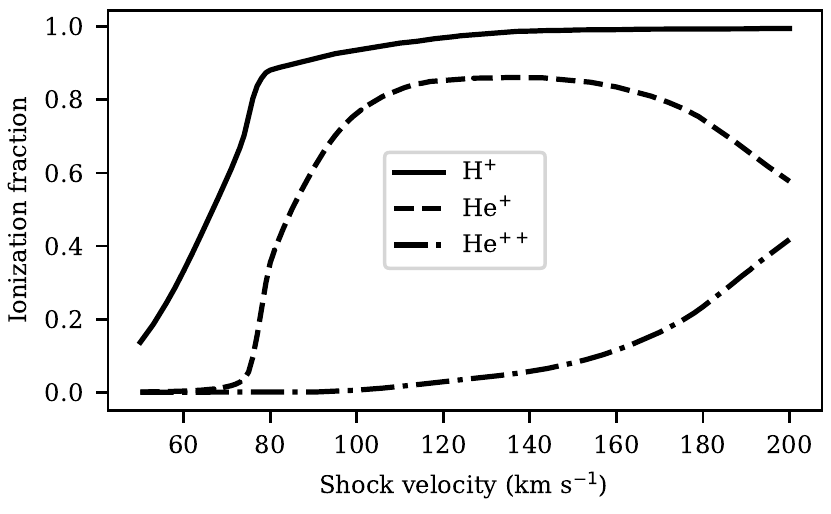}
\caption{Ionization fraction of H$^{+}$, He$^{+}$ and He$^{++}$ versus the shock velocity used as pre-ionization for all shock models computed and shown in this paper.}
\label{fig:preionization}
\end{figure}

The ionization state of the precursor gas is a strong function of shock velocity and its value greatly affects the detailed structure of the shock, hence the prediction of the emissivities as mentioned by \cite{Raymond:1979}. Since this is not a parameter that can be determined directly from our observations, we assume that the shocks present in IC 443 are fast enough to ionise the gas in front of them and no external source other than shocks exists. The computation of the pre-ionization was calculated self-consistently by \textsc{mappings v} as described in \citep{Sutherland:2017}. The pre-ionization was evaluated only once using solar abundance for shock between 50 and 200 \kms with pre-shock density of 20 \cmc and a transverse magnetic field of 1 \muG. The result is shown in Fig.\ref{fig:preionization}. Since the pre-ionization changes very little under similar shock parameters, we have re-used the same pre-ionization values function of the shock velocity for all models computed in this paper. Preventing the evaluation of the pre-ionization for each model allows faster computation, therefore resulting in virtually the same results concerning emissivity predictions. The models presented in this paper do not include the emission emanating from the precursor gas since the shock velocities are not fast enough (< 150 \kms) to produce any appreciable emission, as mentioned in \cite{Dopita:1996}.

\input{table2.tex}

\subsubsection{Line intensity evolution}
\label{sec:emission_evolution}

\begin{figure}
\includegraphics{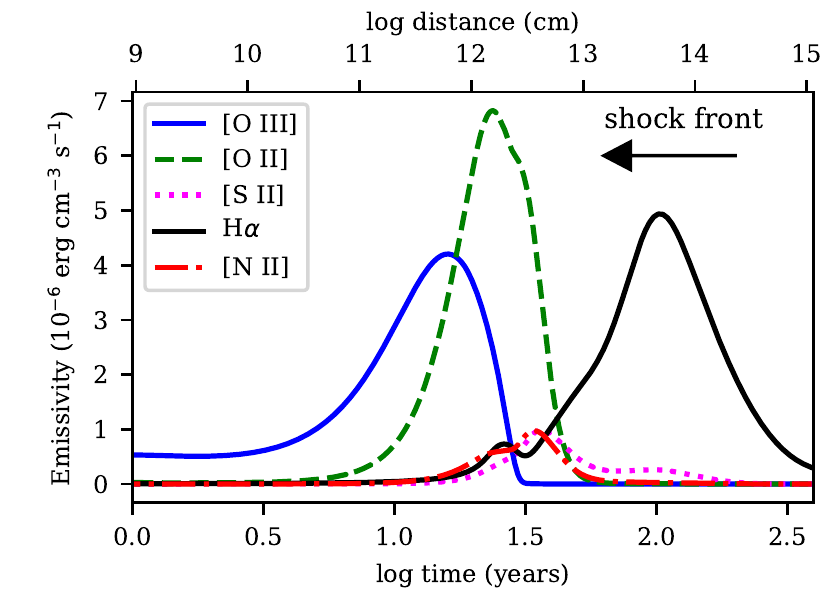}
\caption{Emissivity profile of \OIII $\lambda$5007, \OII $\lambda\lambda$3726+3729, \NII $\lambda$6583, \SII $\lambda\lambda$6716+6731 and \Ha versus the distance (bottom) and the distance (top) for a 100 \kms shock propagating into a density of n$_{0}$ = 20 \cmc and a magnetic field of B$_{0}$ = 1 \muG. The abundances used are given Table \ref{tab:abundances}.}
\label{fig:emiss_profile}
\end{figure}

Fast shocks are effective at ionizing the gas; the faster the shock, the larger the volume of gas is ionized and the longer it will take to recombine. The recombination time depends on several parameters such as the shock velocity, the pre-shock density and the abundances. Whatever the set of parameters, the highest ionization species recombine first, followed by the lowest. This means that the emission lines intensities vary with time. The rate at which the recombination occurs depends on many complex processes and is determined by shock modelling codes such as \textsc{mappings}. As an example, Fig. \ref{fig:emiss_profile} shows the emission line intensity variation for \OIII $\lambda$5007, \OII $\lambda\lambda$3727+3729, \NII $\lambda$6584, \SII $\lambda\lambda$6716+6731 and \Ha for a 100 \kms shock propagating into a region with an electron density of 20 \cmc . It can be seen that \OIII recombines before \Ha and at a certain time both emission are superimposed. This time delay thus leads to a variation of the \OIII/\Ha ratio with time. A high \OIII/\Hb is thus a good indicator of a recent shock and can be used as a compass showing the direction of the shock front. Fig. \ref{fig:colormap_OIII_Hb} clearly shows this phenomenon, where the \OIII/\Hb ratio indicates the regions where shocks are the most recent and at the same time indicates the direction of the blast wave. The same happens with the \OII/\Ha ratio (Fig. \ref{fig:colormap_OII_Ha}). This effect is far less obvious in our observations with the \NII/\Ha and \SII/\Ha ratios since the physical separation between \NII, \SII and \Ha are not as large as those for  \OIII and \OII. 

It must be mentioned that the emission line structure generated behind a shock can barely be observed from our observations considering their spatial resolution. At the distance of IC 443, estimated to be 1.9 kpc \citep{Ambrocio-Cruz:2017}, our seeing-limited spatial resolution of 1$^{\prime\prime}$ corresponds to a spatial distance of 2.8$\times$10$^{14}$ cm. With the example given in Fig. \ref{fig:emiss_profile} and for similar shock parameters, nearly all of the observed emission lines are located within a single pixel in our intensity and ratios maps. Therefore, we suppose that the emission coming from one pixel can be interpreted with the help of one shock model in which the total emissivities can be summed. 

The parameter used in this paper to compute truncated models of incomplete shocks is the cut-off temperature (T$_{cut}$). It corresponds to the temperature of a region located a certain distance behind the shock at which the computation of a model has been stopped. A model can be considered complete when T$_{cut}$ reaches 1000 K below which no further significant emission in the considered species is produced. Very few truncated models can be found in the literature and were computed to match very precise object such as filaments in the Cygnus Loop \citep{Raymond:1980, Contini:1982, Raymond:1988} and a low-excitation Herbig-Haro object \citep{Binette:1985}. 

The large \OIII/\Hb ratios found in some regions of IC 443 suggest the presence of incomplete shocks and the absence of grids with truncated models re-enforced the need to compute one specially adapted to IC 443, taking in consideration the other shock parameters covered in the next sections. 

\subsubsection{Shock velocity}
\label{sec:shock_velocity}

When only observations in the optical regime are available, the shock velocity can derived from the \OIII $\lambda$5007 to \OII $\lambda$3727 ratio, as shown by \cite{Dopita:1977A} (Fig. 7). His figure can be slightly modified to take into account incomplete shock models. Fig. \ref{fig:emiss_profile} shows that the \OIII and \OII emissivities vary differently with time, therefore influencing the \OIII/\OII ratio at a given moment. Fig. \ref{fig:oiii_oii_ratio} shows the variation of the \OIII/\OII ratio as a function of the shock velocity and fixed cut-off temperature during model evaluation. Two important points can be made from this figure: 

\begin{figure}
\includegraphics{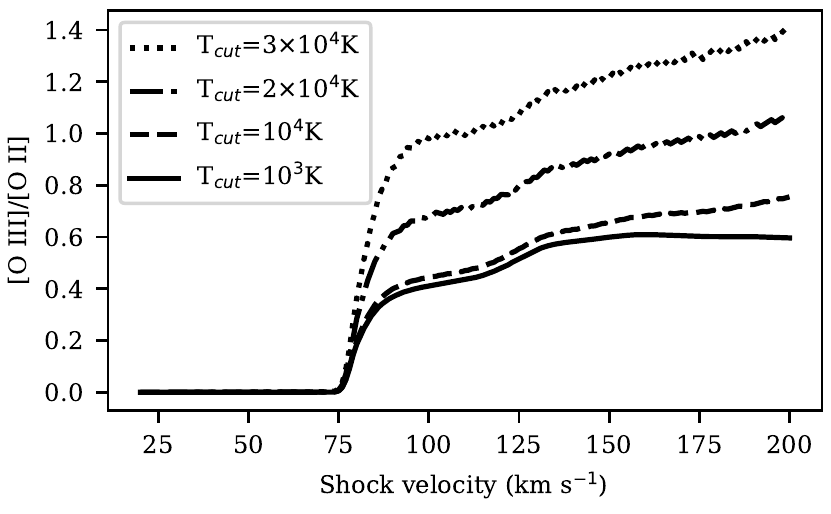}
\caption{Variation of the \OIII $\lambda$5007/\OII $\lambda\lambda$3726+3729 ratio as a function of the shock velocity and cut-off temperatures for n$_{0}$ = 20 \cmc and B$_{0}$ = 1\muG. The abundances used are given Table \ref{tab:abundances}.}
\label{fig:oiii_oii_ratio}
\end{figure}

\begin{itemize}
\item[--] The \OIII/\OII  ratio becomes useful at a shock velocity around 75 \kms,coinciding with the emergence of \OIII $\lambda\lambda$4959,5007 emission. It has to be noted that the velocity at which \OIII appears depend on the ionization state of the precursor gas during model evaluation. For example, a shock propagating into a neutral medium requires at least a 100 \kms shock to produce \OIII. In either case, fast shock can be identified directly by observing filaments emitting \OIII, as seen in Fig. \ref{fig:image OIII}. Regions without \OIII, while emitting lower ionization lines such as \Ha, \NII and \SII, indicate the presence of slower shocks (< 75 \kms).  
\end{itemize}

\begin{itemize}
\item[--] Deriving a shock velocity with the \OIII/\OII ratio requires a precise knowledge of the shock age. Even then, when the shock can be considered as complete (T$_{cut}$=10$^{3}$ K), the usefulness of the \OIII/\OII ratio is limited to a shock velocity range between 75 and 200 \kms, faster shocks producing \OIV at the expense of \OIII. Despite these limitations, the \OIII/\OII ratio can still be useful to approximate element abundances and will be discussed in section \ref{sec:abundances_estimate}.
\end{itemize}

\begin{figure}
\includegraphics{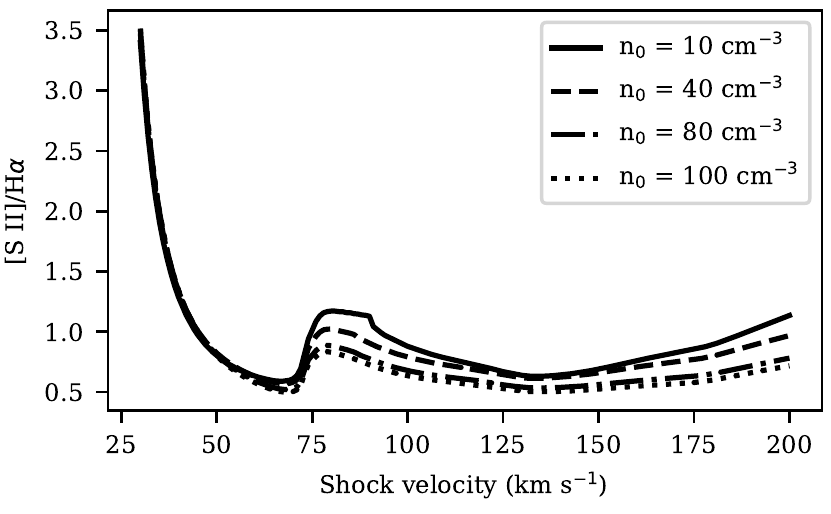}
\caption{Variation of the \SII $\lambda\lambda$6716+6731/ \Ha ratio as a function of the shock velocity and pre-shock densities for B$_{0}$ = 1\muG. The abundances used are given Table \ref{tab:abundances}.}
\label{fig:sii_ha_ratio}
\end{figure}

As mentioned above, the absence of \OIII, along with the presence of lower ionization lines is a good indication of regions being excited by slower shocks. Another indicator that this is the case is the presence of a particularly high \SII/\Ha ratio, as seen in Fig. \ref{fig:colormap_SII_Ha}. Shock models presented in Fig. \ref{fig:sii_ha_ratio} indicate that shocks with a velocity lower than 30 \kms could be responsible for the observed emissions and does not depend on the pre-shock density. 

\begin{figure}
\includegraphics{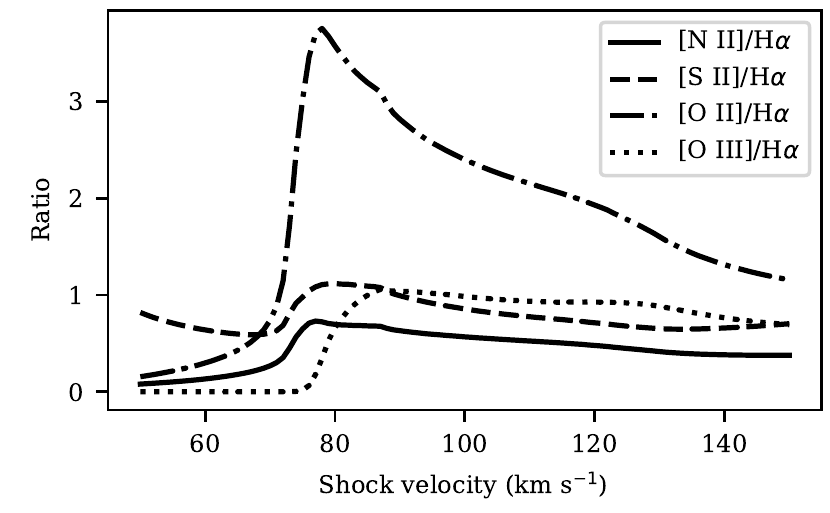}
\caption{Variation of the \OIII/\Ha, \OII/\Ha, \NII/\Ha and \SII/\Ha ratios for shock velocities between 50 and 150 \kms, n$_{0}$ = 20 \cmc and B$_{0}$ = 1 \muG. The abundances used are given Table \ref{tab:abundances}.}
\label{fig:ratio_vs_shock}
\end{figure}

The \NII/\Ha ratio is also linked to the shock velocity as shown in Fig \ref{fig:ratio_vs_shock} alongside with other ratios. We can see from Fig. \ref{fig:colormap_NII_Ha} that the \NII/\Ha ratios are lower in locations where the \SII/\Ha are higher and \OIII absent, therefore, where the shock velocities are the slowest. 

\subsubsection{Pre-shock density}
\label{sec:pre_shock_density}

Electron densities can be found using the \SII $\lambda$6716/$\lambda$6731 \citep{Weedman:1968} which are available from our data. Measurements obtained by others, using \OII and \SII lines, show clear variations across the nebula. Measurements of brights filaments \citep{Osterbrock:1958} and three positions by \citet{Parker:1964} show density fluctuations within the range of 100-600 \cmc with an average of 350 \cmc. \citet{Fesen:1980} found slightly different values, with electron densities ranging from 100 to 400 \cmc, with an average of 200 \cmc. Using infrared observations, \citet{Rho:2001} found an electronic density of 500 \cmc. 

Figure \ref{fig:colormap_SII_SII} shows the variation of the \SII ratio from our observations. The ratio varies from 0.70 to 1.30, with an average of 1.22. Using \textsc{pyneb} \citep{Luridiana:2012,Luridiana:2015} with the atomic data of sulphur taken from the \textsc{chianti} database version 8 \citep{Dere:1997,DelZanna2015}, we calculated the corresponding electronic densities, which translate a minimum and maximum being respectively 106 and 2477 \cmc. These values are in agreements between previous measurements, but ours show regions with significantly higher electronic densities. Most of them are located in the northeastern part of IC 443, but also in some localised filaments as well. This indicates that the blast wave is encountering a higher pre-shock density in those regions.

\begin{figure}
\includegraphics{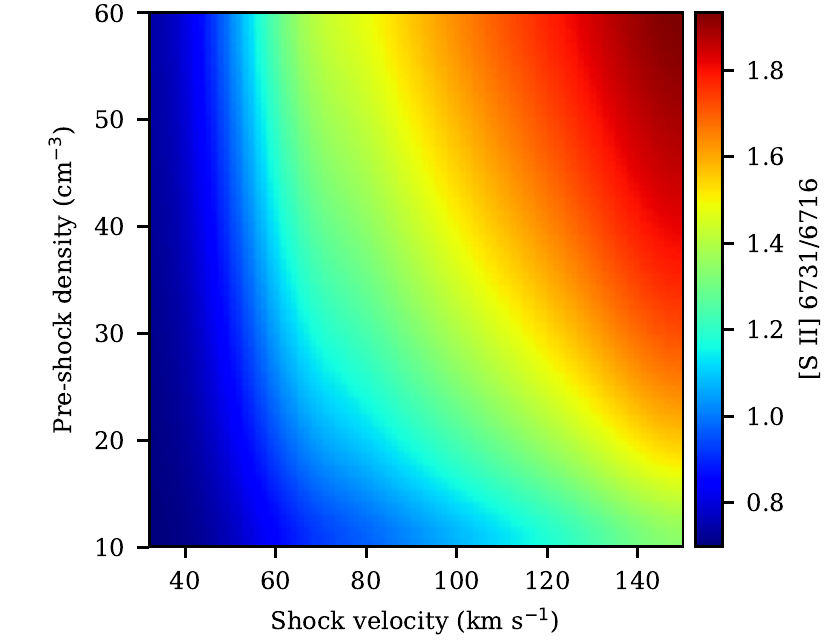}
\caption{Variation of the \SII $\lambda$6731/$\lambda$6716 ratio as a function of the preshock density and the shock velocity for solar abundances and B$_{0}$ = 1 \muG. This figure was made using 16384 individual shock models evaluated with \textsc{mappings}. }
\label{fig:density_estimate}
\end{figure}

Values of the pre-shock density can be estimated with the help of Fig. \ref{fig:density_estimate}. Assuming a constant magnetic field of 1 \muG throughout IC 443, the pre-shock density can be estimated when the shock velocity is known. Using the \SII $\lambda$6716/$\lambda$6731 ratio, and \OIII/\OII ratio to determine the shock velocity, we found that most pre-shock densities vary between 20 and 60 \cmc. These values are higher than those measured by \cite{Fesen:1980}, but this is not surprising since we measured higher electronic densities. We could not measure lower pre-shock densities since the velocity could not be measured at all locations. Measuring the pre-shock density for slow shocks (< 40 \kms) could prove challenging since the \SII $\lambda$6731/$\lambda$6716 ratio then varies very little, as shown in Fig. \ref{fig:density_estimate}. This effect can be seen by comparing the locations of high \SII/\Ha ratio (slow-shocks-Fig. \ref{fig:colormap_SII_Ha}) and the \SII ratio map (Fig. \ref{fig:colormap_SII_SII}), where no clear relation can be seen between \SII/\Ha line and \SII $\lambda$6731/$\lambda$6716 line ratios morphologies. 

\begin{figure}
\includegraphics{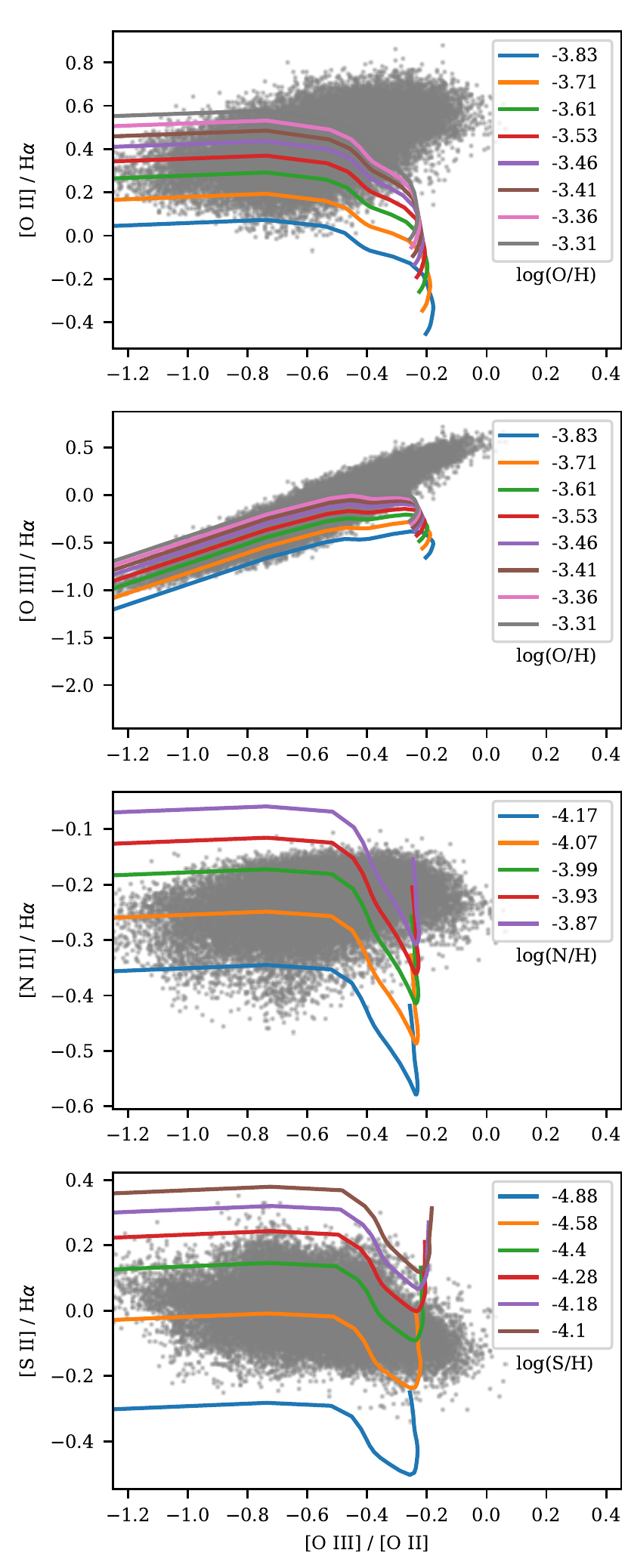}
\caption{Diagnostic diagram of \OII/\Ha, \OIII/\Ha, \NII/\Ha and \SII/\Ha versus \OIII/\OII ratio. The point cloud represent the observations and the colored lines shock models evaluated with different metallicities for shock velocity between 70 \kms and 200 \kms, n$_{0}$ = 20 \cmc and B$_{0}$ = 1 \muG.}
\label{fig:abundances_estimate}
\end{figure}

\subsubsection{Abundance range estimates of O, N and S}
\label{sec:abundances_estimate}

With the emissions of \OIII, \OII, \NII, \SII and \Ha, it is possible to estimate the probable abundances of oxygen, nitrogen and sulphur. In order to achieve this, it is required to estimate the shock velocity, which can be derived from the \OIII/\OII ratio as mentioned in section \ref{sec:shock_velocity}. Inspired by the work of \cite{Hester:1983}, the\OIII/\OII ratio can be used as an abscissa while the \OII/\Ha, \OIII/\Ha, \NII/\Ha and \SII/\Ha can be used as the ordinate. This configuration allows putting a physical meaning on each axis, the abscissa, the shock velocity and the ordinates a mean to adjust the abundances. This convenient setup works only under some circumstances which are going to be explained in detail in this section. 

Fig. \ref{fig:abundances_estimate} shows diagrams using the axes configuration described above, showing both the observations made with SITELLE (grey dots) taken from the ratio maps and complete shock models calculated with different abundances (coloured lines). We used a fixed pre-shock density of 20 \cmc and a magnetic field of 1 \muG for all models. It has to be noted that a different pre-shock density would have little effect on the models shown in Fig. \ref{fig:abundances_estimate} as long as the selected value remains inside the given pre-shock density range measured in IC 443 from the \SII ratio in section \ref{sec:pre_shock_density}. The shock velocity chosen run from 70 \kms to 200 \kms varying from left to right on the diagram. By selecting the \OIII/\OII ratio, only fast shocks were selected at the expense of slower shocks. Therefore, the data points do not include all the observed spectra with SITELLE. 

An interesting feature coming from these diagrams is how well the slope on the \OIII/\Ha diagram is reproduced, clearly showing the relation between the intensity of \OIII and the shock velocity. On the contrary, the slopes on the other diagrams (\OII/\Ha, \NII/\Ha and \SII/\Ha) are very low, meaning that these ratios are affected very little inside a narrow shock velocity range ($\approx$ 80 to 100 \kms). This feature can be exploited in order to grossly estimate plausible abundance ranges of O, N and S, since the \OII/\Ha, \NII/\Ha and \SII/\Ha vary linearly with abundances change. This method is by no means to be used to determine precisely the abundances in IC 443. The variation of \OII/\Ha, \NII/\Ha and \SII/\Ha are most likely the result of an intricate combination of several other physical parameters and will be discussed in section \ref{sec:bi_shock_models}. What can be said with reasonable certainty with these diagrams is that some abundances could be discarded. For example, an abundance  of nitrogen with log(N/H)=-4.17 seems to be a little high; as for sulphur,  log(S/H)=-4.88 seems a little too low for the shock velocity range considered. As for oxygen, it seems that a wide range of abundances could explain the observed emissions. 

The derived abundances range derived from Fig. \ref{fig:abundances_estimate} are somewhat close the solar values determined by \cite{Asplund:2009} with the exception of nitrogen and sulphur being too low with values evaluated at log(N/H) = -4.17 and log(S/H) = -4.88. As for the solar abundance of oxygen, its derived value of log(O/H) = -3.31 could explain the observed spectrum. The abundances adopted in this paper from helium to zinc (excepting N and S) were taken from the compiled solar abundances of \cite{Asplund:2009}, \cite{Scott:2015a}, \cite{Scott:2015b} and \cite{Grevesse:2015}. The abundance of nitrogen and sulphur were chosen from Fig. \ref{fig:abundances_estimate} fitting the model on the point clouds and the list of abundances used to compute all presented models is given in Table \ref{tab:abundances}.

\subsection{Bi-shock models with multiple parameters}
\label{sec:bi_shock_models}

Previous sections have demonstrated that many shock parameters are responsible for the observed spectra in IC 443. So far in this paper, as well as in previous studies, the observations were mainly compared with single shock models that include only one set of parameters. While this rather simplistic technique allows to derive important parameters, such as the shock velocity, pre-shock density and even approximate the abundances, it does not always explain all the observed spectra. Fig. \ref{fig:abundances_estimate} exposes several flaws when it comes to compare shock models with observations too simplistically. The most obvious one is the inability to reproduce all observations above log(\OIII/\OII)$\approx$-0.22. The presence of elevated ratios can be explained by the presence of incomplete shocks, as mentioned in section \ref{sec:shock_velocity} (Fig. \ref{fig:oiii_oii_ratio}), which demonstrates the sensitivity of the \OIII/\OII ratio to shock completeness. However, it does not suffice to run a grid of truncated models in order to improve the match between the models and observations. Fig. \ref{fig:abundances_estimate} shows that several shock models with different abundances allow the overlapping of models over observations. While these diagrams allow approximating the plausible abundances in IC 443, it does not explain the reason for the point dispersion and therefore the origin of the great variety of line ratios observed in IC 443. The variation of abundances alone would not make sense. Ideally, each spectrum should be represented by one model with its very own set of parameters. 

The filamentary structure in IC 443 is pretty intricate with many filaments criss-crossing, as viewed from the observer's point of view. Upon close inspection, many distinct interlaced filaments show different intensities for a given emission line. As we show in section \ref{sec:kinematics}, the receding and approaching components of the expanding bubble show very different morphological structures, suggesting that shock conditions on a given line of sight could be different and that the summation of both components on the same line of sight can lead to a spectrum that cannot be interpreted with the help of a single shock model. Therefore, many locations obviously required more than one shock model. 

In an attempt to better reconcile models and observations, we have computed a grid of 57,032 shock models with the parameters and intervals listed in Table \ref{tab:shock_parameters}. The shock velocity range was chosen in order to reflect the measurements presented in section \ref{sec:shock_velocity} and the pre-shock densities with those derived in section \ref{sec:pre_shock_density}. We assume that they are shocks with different ages, as mentioned in section \ref{sec:emission_evolution}. We also assume that the abundances do not vary significantly in the observed region and selected one set of abundances as determined in section \ref{sec:abundances_estimate}. 

\input{table3.tex}

To simulate the effect of multiple filaments on the same line of sight, we have to sum a given number of distinct shock models according to the number of components. Since we do not always have a clear idea about the exact shock parameters for each component, we can generate a new grid of models based on another grid using permutations. The total number of permutations of such shock models is given by the following equation:

\begin{figure*}
\includegraphics{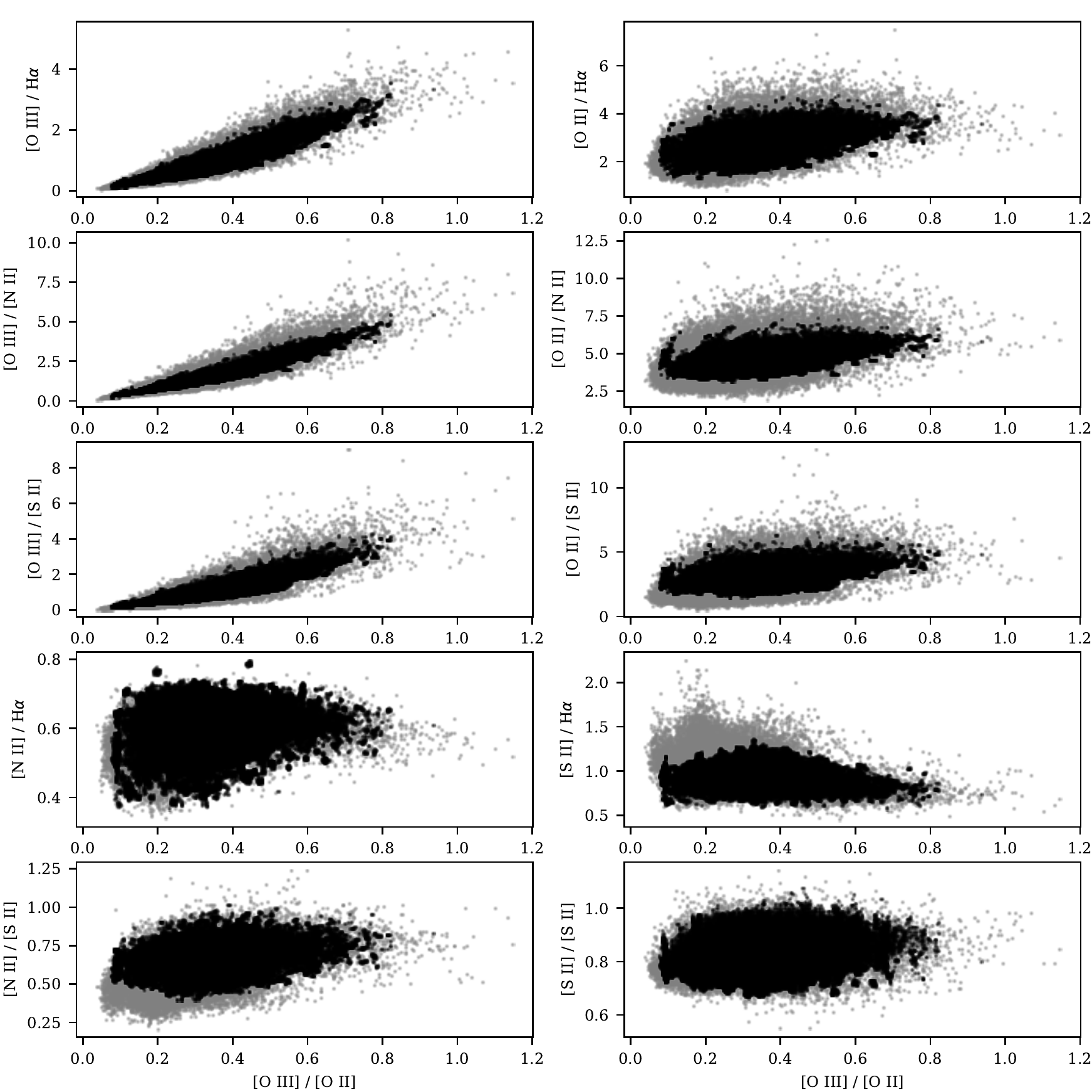}
\caption{Bi-shock models in black on top of observations in gray. The combination of shock parameters for the shock velocity, cut-off temperature and pre-shock density are shown in a histogram in Fig.\ref{fig:histograms}. The abundances used are shown in Table \ref{tab:abundances}}
\label{fig:bimodels}
\end{figure*}

\begin{equation}
C\left ( n,r \right )=\binom{n}{r}=\frac{\left ( r+n-1 \right )!}{r!\left (n-1  \right )!}
\label{eq:permu}
\end{equation}

where $n$ is the total number of shock models available in a grid and $r$ the number of models to be summed, or in other words the number of components on the same line of sight. Fabry-Perot observations conducted by \cite{Ambrocio-Cruz:2017} have indeed shown that up to four velocity components can be found along the line of sight at some positions in the nebula. 

From equation \ref{eq:permu}, we see that the total number of permutations can become very large when $r$ is significantly larger than 2, depending on the total number of shock models available in a grid. Since our observations do not allow us to determine the exact number of components at every point in IC 443 and due to computational limitations, we choose a value of $r=2$. This leads to a total number of of 1,626,353,028 models permutations which were systematically compared with every spectrum shown in Fig. \ref{fig:bimodels} (gray points) made from 35,479 spectra taken from our ratio maps. In total, 5.3$\times$10$^{13}$ comparisons where made and models who predicted all 10 line ratios presented in Fig. \ref{fig:bimodels} with a margin of error smaller than 5\% were kept. The result is given on the same figure (black points) which shows 16,215 models overlapping the observations. About 45 \% of the selected spectra where fitted by our models. Most of the spectra can be represented by more then one model with different parameters leading to degeneracies. 

\begin{figure}
\includegraphics{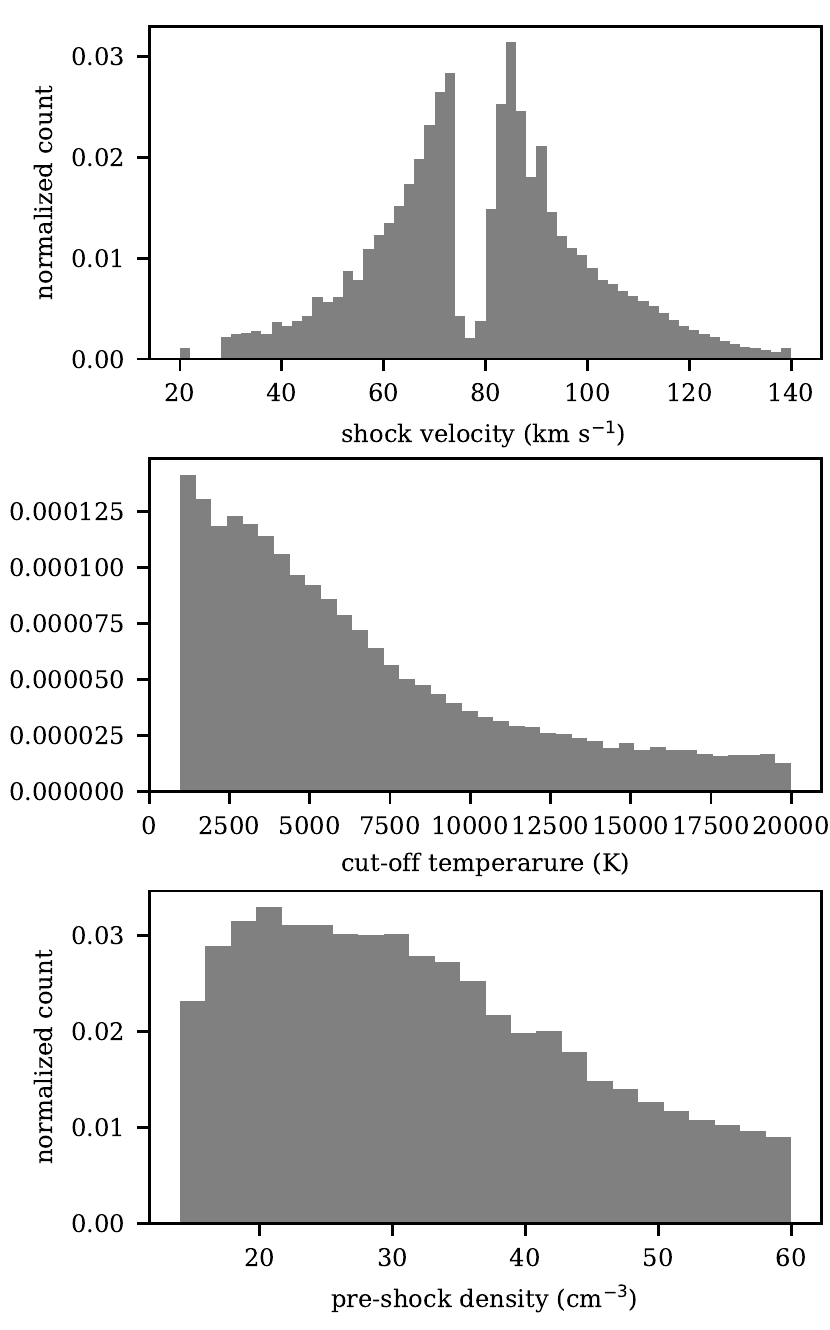}
\caption{Histogram showing the distribution of the combination of two models for the shock velocity (top), cut-off temperature (middle) and pre-shock density (bottom) shown in Figure \ref{fig:bimodels}.}. 
\label{fig:histograms}
\end{figure}

Fig. \ref{fig:histograms} shows the frequency range of parameters that best fit the observations. These diagrams were made including the degeneracies. From that figure, we see that the spectra could be explained by the presences of shocks velocities range between 20 and 140 \kms with a clear preference toward velocity around 75 \kms. Interestingly, there is a dip around that velocity where there should be a peak. This can be explained by the fact that the \OIII/\OII ratio quite steeply above 75 \kms as shown in Fig. \ref{fig:oiii_oii_ratio}. Any small error on the \OIII/\OII ratio can easily "shift" the velocity on the other side of the dip as shown in Fig. \ref{fig:histograms}. The same figure also shows a variety of cut-off temperatures, and thus shocks with different ages, favoring complete shocks.  Models tend to favour pre-shock densities around 20 \cmc but lower densities could not be excluded. 

The most important result from Fig. \ref{fig:bimodels} is that the combination of two shock models allows us to reproduce some observed spectra with log(\OIII/\OII)$\geq$0.60, which were could not be explained with a simple model. This suggests that the individual spectra are indeed the combination of at least two components along the line of sight. Despite this success, some observation could not be reproduced. For example, there is a clear gap in the upper left for the \SII/\Ha diagram and the lower left for the \NII/\SII diagram in Fig. \ref{fig:bimodels} for which none of our summed models were able to fit the observations. However, single shock models can easily reproduce the observed ratios. It is probably the case with many other spectra for which one or more than three components can potentially provided a better fit to the spectra. However, more components will undeniably bring more degeneracies at which point the shock parameters found might lose their meaning and cannot provide additional informations. 

It seems that one set of abundances is sufficient in order to reproduce the line intensities dispersion. In order words, that is no need to vary the abundances for specific spectrum in order to better fit the observations. We tested this with different abundances of O, N and S around the ranges that can be inferred from Fig. \ref{sec:abundances_estimate} and saw no significant improvements. However, we noticed that an abundance of oxygen closer to solar value was preferable in order to better fit the observations.

\begin{figure}
\includegraphics{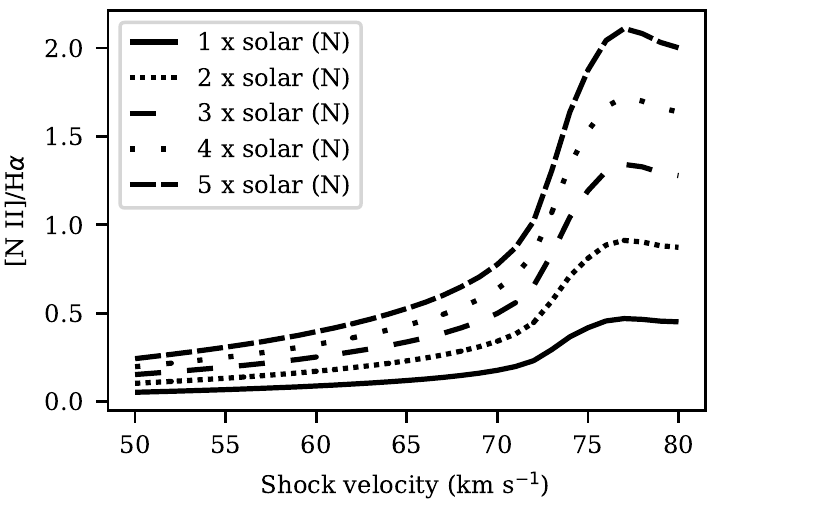}
\caption{Variation of the \NII/\Ha ratio for different nitrogen abundances vs the shock velocity. 1$\times$solar corresponds to a nitrogen abundance of log(N/H) = -4.17. The other abundances used to compute the curves are given in table \ref{tab:abundances}}. 
\label{fig:overabundN}
\end{figure}

We notice however high \NII/\Ha ratios (Fig. \ref{fig:colormap_NII_Ha} south of the field of view) which might suggest a region enriched with nitrogen. Since the \OIII/\OII ratio cannot be obtained for these locations, we were unable to estimate a shock velocity and therefore accurately determine the nitrogen abundance. However, the absence of \OIII for most of the location where \NII/\Ha is elevated suggests shocks slower than 75 \kms. In that case, models calculated with a nitrogen abundance close to solar are unable to reproduce the observed \NII/\Ha and higher abundance must be considered. Figure \ref{fig:overabundN} shows the variation of the \NII/\Ha ratios in function of different nitrogen abundances. It can been seen that for given \NII/\Ha ratio an accurate determination of the abundances is very sensitive to the shock velocity. When the velocity is unknown, multiple plausible abundances could be responsible for a unique ratio.

From figure \ref{fig:colormap_NII_Ha} we measured a ratio of \NII/\Ha as high as 1.2. From figure \ref{fig:overabundN} we can see that a shock velocity around 75 \kms with an abundance of nitrogen of 3 times the solar could explain the observation. However, if the shock velocity is slower, a higher abundances of nitrogen must be considered.

The location of this enrichment in nitrogen is interesting since it is located toward the center of the nebula and might be ejecta from the progenitor which was blown outward by its stellar winds prior to its explosion. 

\section{Kinematics}
\label{sec:kinematics}

A few measurements of the radial velocities across IC 443 have been obtained using long-slit spectroscopy \citep{Lozinskaya:1969b} and Fabry-Perot interferometer \citep{Lozinskaya:1969a, Pismis:1974, Lozinskaia:1979, Ambrocio-Cruz:2017}. The radial velocities measured by the various observations reported velocities ranging from approximately $-180$ to $+135$ \kms and vary quite significantly across the nebula. 

While our primary intent was not to study the kinematics (hence the relatively low spectral resolution of our cubes: 130 km/s/channel at \Ha), the high signal-to-noise of the SN3 cube, as well as the relatively similar intensity of four emission lines within the limits of this filter have allowed us to separate the ionized gas into two distinct and morphologically very different components. Two approaches, leading to very similar results, were followed.

\begin{figure*}
\includegraphics[scale=0.45]{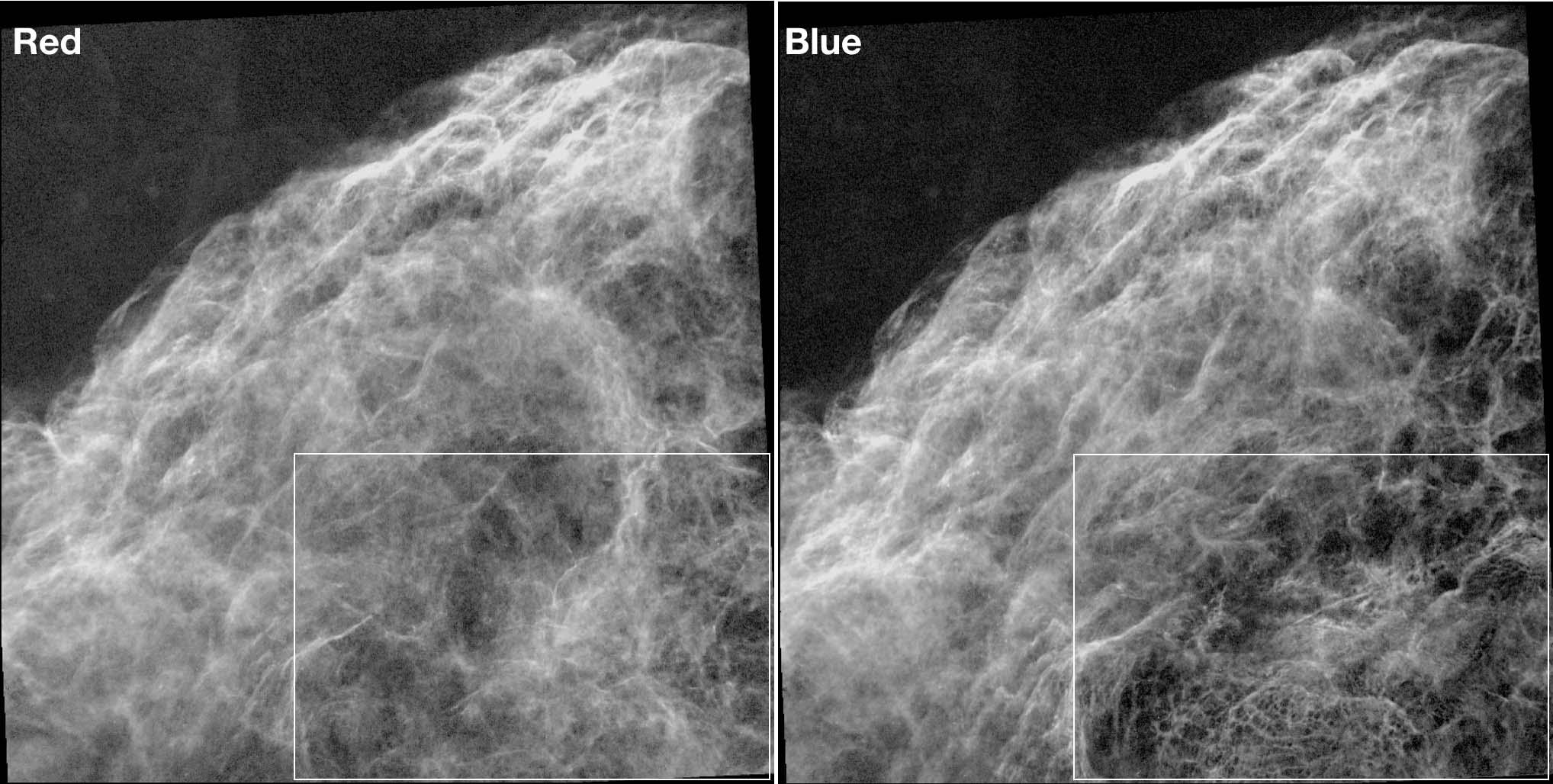}
\caption{Red and blue components of all emission lines present in the SN3 filter, obtained by selecting the frames in the data cube corresponding
to the wings of the lines (see text). The region which was analyzed kinematically with ORCS is indicated by a white rectangle.}
\label{fig:blue-red}
\end{figure*}

\begin{figure}
\includegraphics[width=\columnwidth]{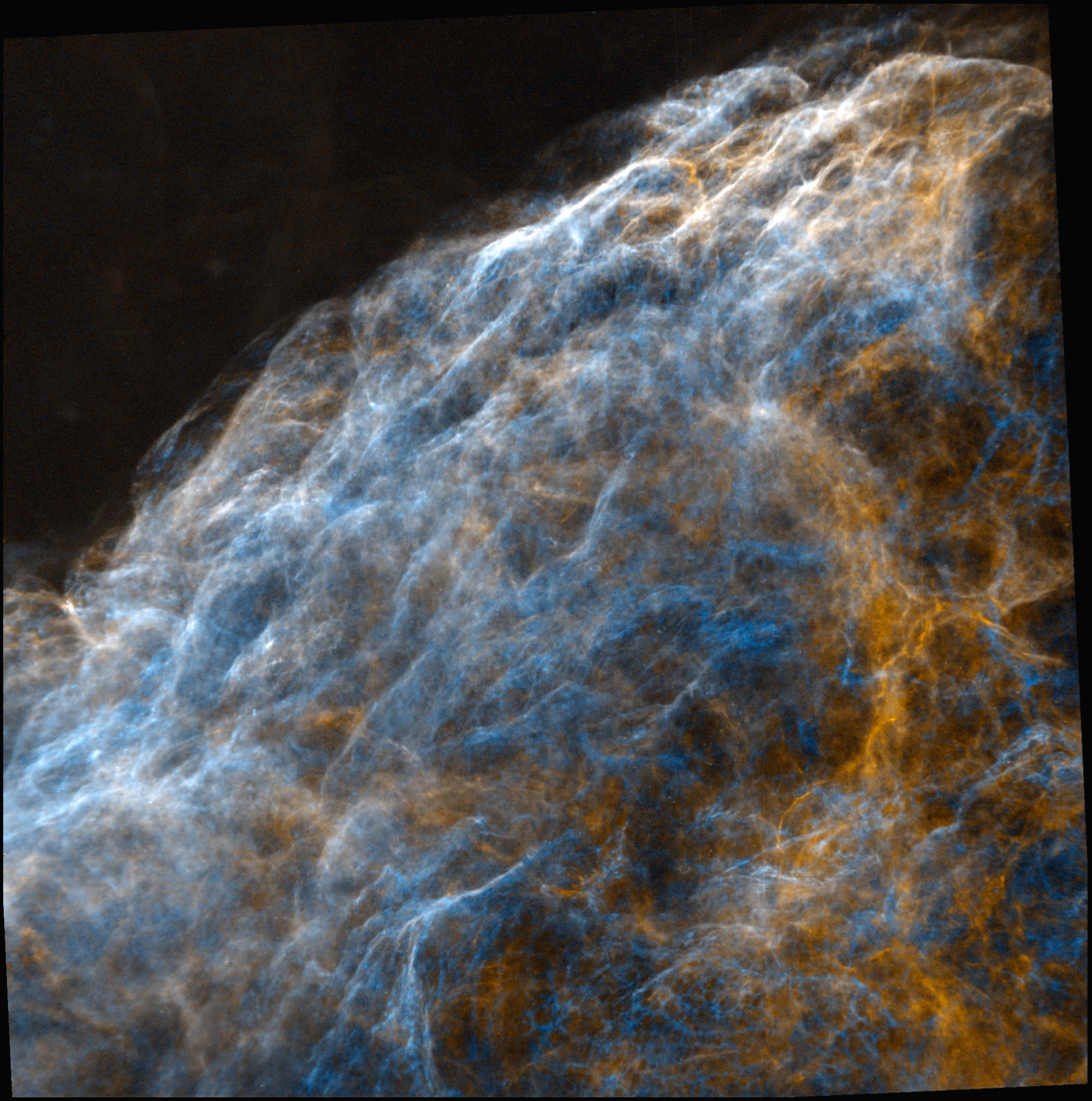}
\caption{Color-coded combination of the two images shown in Fig. \ref{fig:blue-red}.}
\label{fig:blue-red-color}
\end{figure}

The first one is very simple: the spectral sampling and the expansion velocity of the nebula are such that emission from the nebula is visible on three consecutive frames of the data cube for each line: \Ha-red, \Ha-green and H$\alpha$-blue; \NII $\lambda 6584$-red, and so on. The line centroid (and hence maximum emission) does not always coincide with the middle frame and the balance between the red and the blue components is not identical from one line to the next. However, combining all the red frames together and doing the same for all the blue frames results in two emission-line maps sampling velocities separated by 260 \kms. These images, which essentially sample the blue and red wings of all emission lines, are shown in Fig. \ref{fig:blue-red}, and a color composite is presented in Fig. \ref{fig:blue-red-color} . While the upper left part of the nebula in this field looks pretty similar in the red and blue channels, the red and blue components of the lower right part, closer to the center of the supernova remnant and thus where the Doppler shift is expected to be larger, are dramatically different. While this method is relatively straightforward and provides a good glimpse at the three-dimensional structure of the nebula, it does not offer a quantitative distribution of the velocities at play.

\begin{figure}
\includegraphics{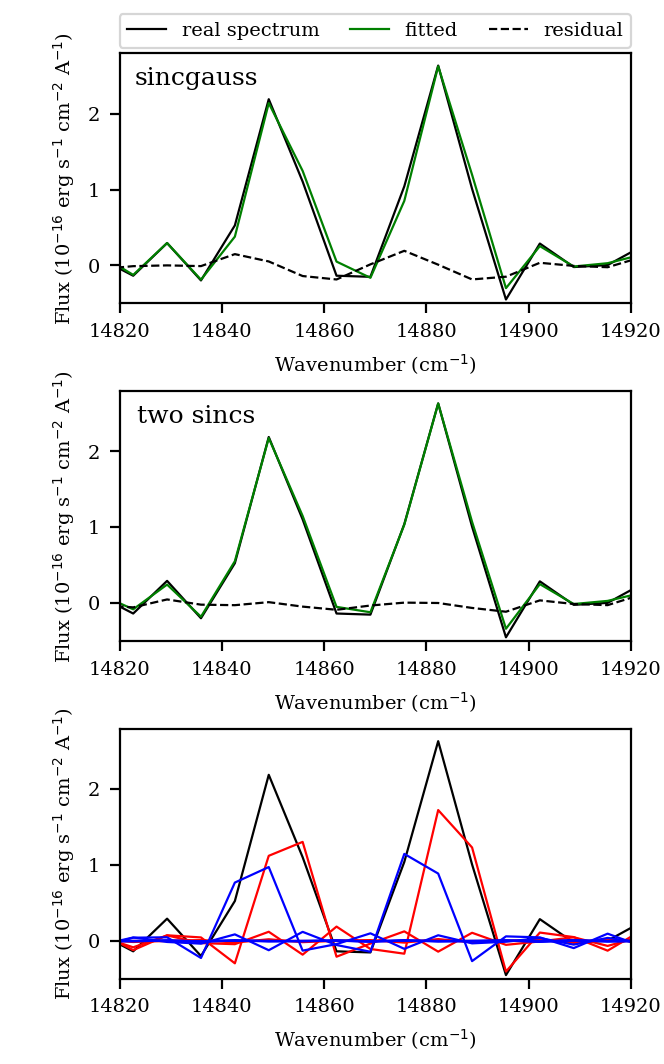}
\caption{Example of the fitting functions used by ORCS, for the \SII 6717,31 doublet in a single pixel. On the top left panel, a sincgauss function was tried, returning a velocity dispersion, $\sigma = 120$ km/s. On the top right, the result of two sinc functions. The lower panel shows the two components.}
\label{fig:blue-red-fit}
\end{figure}

\begin{figure}
\includegraphics[width=\columnwidth]{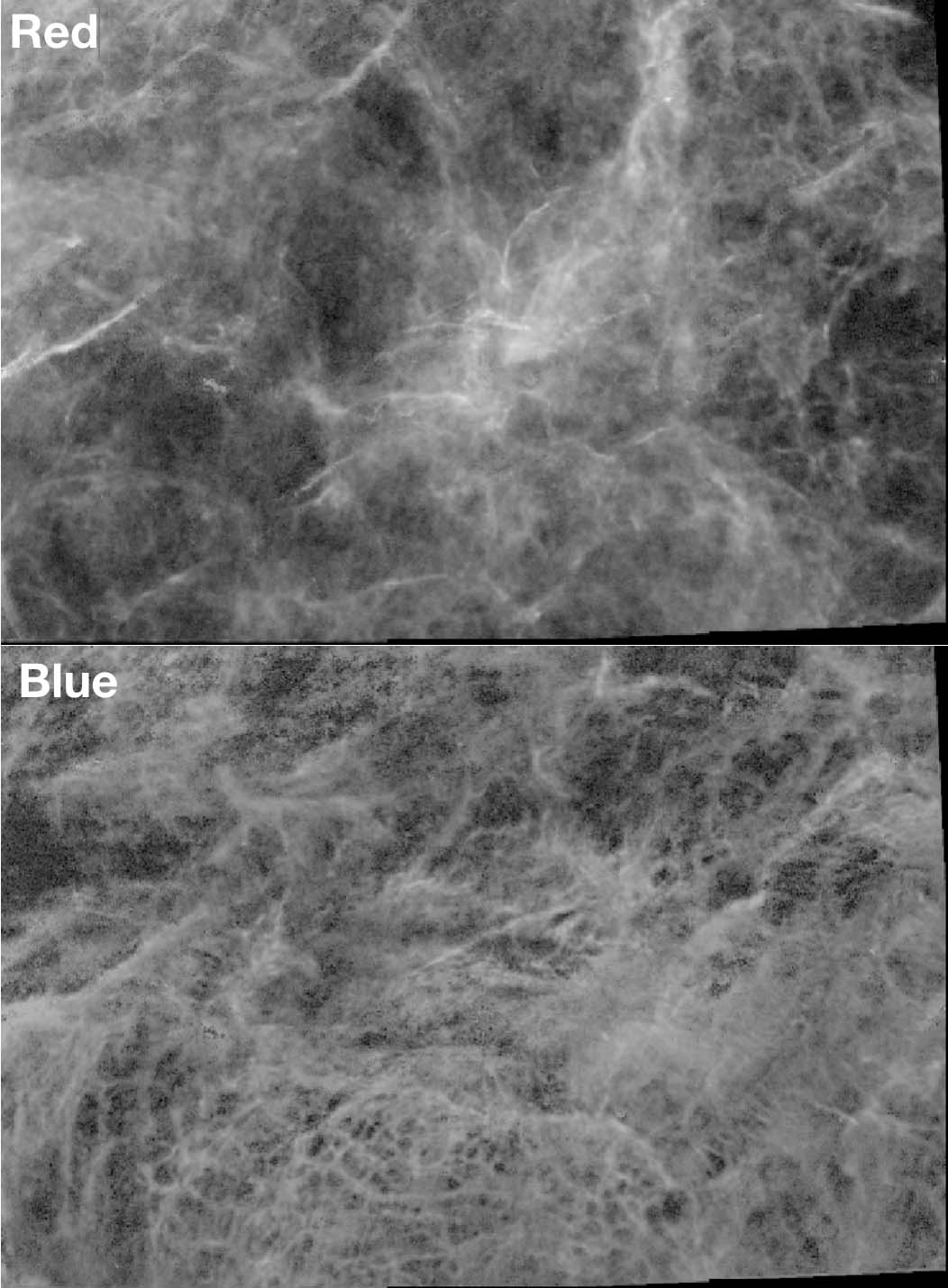}
\caption{Red and blue components in the SW region ($7.3' \times 4.5'$) of the SITELLE field of view, as extracted with ORCS (see text). The similarity between these images and Fig. \ref{fig:blue-red} is striking. The limitations of this method are however visible in the upper left corner of the blue component image, where low-amplitude fringes can be seen.}
\label{fig:blue-red-zoom}
\end{figure}

\begin{figure}
\includegraphics[width=\columnwidth]{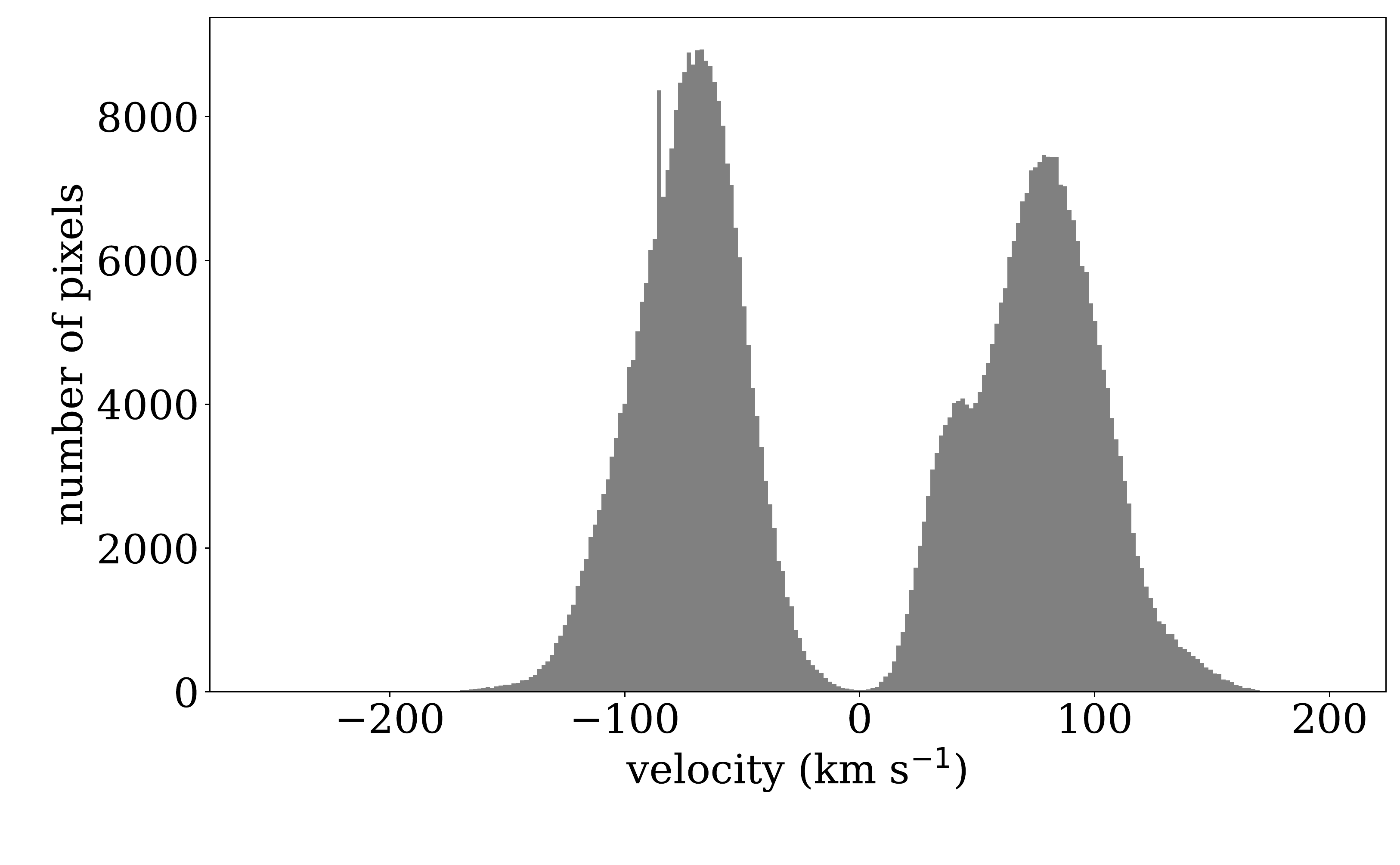}
\caption{Histogram of the velocities obtained with ORCS in the region shown in Fig. \ref{fig:blue-red-zoom}.}
\label{fig:histogram}
\end{figure}

We have therefore used ORCS to quantify this. Among the fitting procedures offered by ORCS, two are particularly adapted to the IC 443 data: as a first approach, the lines can be fitted using a sincgauss function \citep{Martin:2016}, which is a convolution of the instrument line shape -- a sinc -- with a gaussian function characterizing the Doppler broadening of the line; or they can be fitted with two sinc components attributable to two distinct (approaching and receding) components. As expected, the sincgauss fit revealed a clear difference between the upper section of the field, where the sincgauss fit was not significantly different from a single sinc fit (given the low spectral resolution of our cube), and the lower right part, where a significant broadening -- between 75 and 200 \kms -- was clearly detected. We have thus targeted a $7.3 \times 4.5$ arcmin region in the lower right corner of the field (see Fig. \ref{fig:blue-red}) to perform a two-component sinc fit. While ORCS allows to attribute different velocities to the individual lines, we chose to force all the lines present in the SN3 filter to share the same velocity. An example of these two fits on the spectrum of a single pixel is presented in Fig. \ref{fig:blue-red-fit}, and the maps resulting from individual fits to the entire region are shown in Fig. \ref{fig:blue-red-zoom}. The result is strikingly similar to that using the simple approach described above. It is important to note that the limitations of this fitting technique become apparent in the upper left section of the blue component, where very subtle, but nevertheless present, stripes are visible. In this region, the width of the sincgauss function approaches 50 \kms, one fourth of the instrumental line function's width. An histogram of the velocities resulting from the fits is presented in Fig. \ref{fig:histogram}: the two components are clearly separated, by an average of 150 \kms. While the receding component of the nebula presents a simple distribution, the approaching one is clearly bi-modal. There is also a very obvious difference in the morphologies of the two components (Fig. \ref{fig:blue-red-zoom}): the approaching side (very likely the front of the nebula, closer to us) presents small-scale honeycomb-like structures, while the other side (presumably the back, more distant) is smoother and more regular. This dichotomy very likely reflects the structure of the interstellar medium in which the blast wave is expanding (such as the pre-shock density). We also note that the combined flux of all the SN3 emission lines is twice as large in the red component as in the blue one.

While ORCS also provides the flux in the individual lines for both components, we do not wish to over-interpret our rather low spectral-resolution data at this time, but note that a cube with a higher spectral resolution could provide a means to determine the physical properties of both components individually. 
In the previous sections, we have demonstrated that a single shock model could not reproduce the observed line ratios. The different morphologies of the red and blue components indeed suggest a dichotomy in their shock properties.

\section{Concluding remarks}
\label{sec:remarks}

We used the optical imaging Fourier transform spectrometer SITELLE to obtain multi-spectral cubes of the north-eastern part of IC 443. From these cubes, we generated maps allowing the investigation of the spatial variations in spectral line intensity of seven emission lines: \Ha, \Hb, \OII $\lambda$3726+3729, \OIII $\lambda$5007, \NII $\lambda$6583 and \SII $\lambda$6716,6731. We compared the observations with shock models computed with  \textsc{mappings} and estimates important physical properties found in IC 443. Specific results are as follows:

\begin{enumerate}
 	 \item The reddening has been measured from the \Ha/\Hb ratio for the observed region and found to vary significantly at its surface with E(B-V) of 0.8-1.1. 
 	 
	\item The electron density observed in the filaments ranges from less than 100 to 2500 \cmc with higher densities found at the extremity of the north-eastern part of IC 443. 
 	 
 	\item The line intensities vary quite significantly across the field of view  and can be traced to the presence of different shock velocities at different stage of recombination. High \OIII/\Hb and \OII/\Hb ratios are higher towards the direction where the shock front is propagating, due to the presence of incomplete shocks. High \SII/\Ha with no emissions in \OIII toward the center of the observed region indicates the location of shocks with velocity slower then 40 \kms. 
  	
	\item Shock models with a unique set of parameters are unable to reproduce the great majority of the observed emissions. The summation of two shock models (bi-models) with different shock parameters allow a better prediction of the variability of the observed emission line ratios. Bi-models comparisons allow to constraint the probable shock condition inside IC 443 for the observed region. The most probable shock velocities varying from 20 to 150 \kms with 75 \kms being the most common velocity found, the pre-shock density between 20-60 \cmc favouring lower densities and shocks at different stages of recombination as suggested by wide range of the \OIII/\Hb ratios.  
	
	\item A single set of abundances close to solar values combined with varying shock velocity, pre-shock density and shocks with different ages are sufficient to explained the measured ratios. 

	\item Despite the modest spectral resolution of the SN3 cube, we were able to separate the approaching and receding components of the expanding nebula. These two components harbor very different morphological structures, probably resulting from different properties of the interstellar medium in which the blast wave propagates. Higher spectral resolution data cubes across the visible range with SITELLE would allow us to better probe the physical condition on both sides of the nebula.
	
\end{enumerate}

\section*{Acknowledgements}
Based  on  observations  obtained  with  SITELLE,  a  joint project  of  Universit\'e  Laval,  ABB,  Universit\'e  de  Montr\'eal,
and  the  Canada-France-Hawaii  Telescope  (CFHT)  which is  operated  by  the  National  Research  Council   
of Canada,  the  Institut  National  des  Sciences  de  l'Univers  of the  Centre  National  de  la  Recherche  Scientifique 
of France, and the University of Hawaii. The authors wish to recognize and acknowledge the very significant
cultural role that the summit of Mauna Kea has always had within
the indigenous Hawaiian community. We are most grateful to have
the opportunity to conduct observations from this mountain.
AA thanks CONACyT-CB2015-254132 project for supporting his postdoc position at IA-UNAM. Models were performed using computers from projects DGAPA/PAPIIT-107215 and CONACyT-CB2015-254132.
LD is grateful to the Natural Sciences and Engineering Research Council of
Canada, the Fonds de Recherche du Qu\'ebec, 
and the Canada Foundation for Innovation for funding. LD is very grateful to the Department of Physics and Astronomy, University of Hawaii Hilo, as well as the CFHT, for having provided him with an ideal and very enjoyable working environment during his sabbatical stay on the Big Island.
 
All computations from data reduction to analysis were conducted using open source software : \textsc{python} programming language and the GNU Data Language. Special thanks to the people who developed and maintain vital libraries used in this work: \textsc{numpy} \& \textsc{scipy} \citep{Walt:2011}, \textsc{cython} \citep{Behnel:2011} and \textsc{matplotlib} \citep{Hunter:2007}. This work also made extensive use of the \textsc{gnu parallel} shell tool \citep{Tange:2011a}. 




\bibliographystyle{mnras}
\bibliography{bibtex}




\appendix

\section{Figures}
\label{sec:figures}

\begin{figure*}
\includegraphics[width=0.60\textwidth]{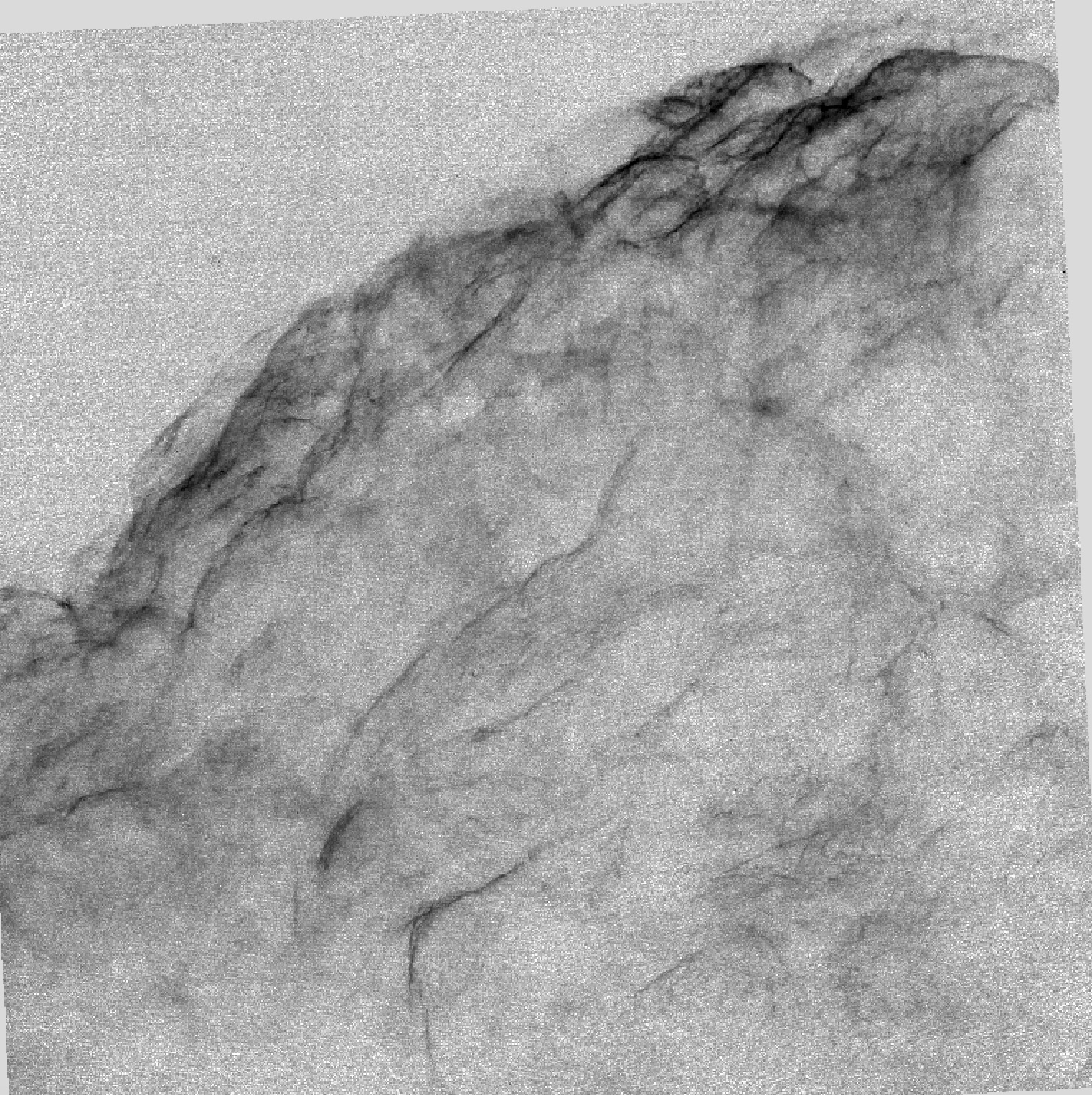}
\caption{The observed region in \OII $\lambda$3727+$\lambda$3729. FOV is 11$' \times 11'$ with North at the top and East to the left.}
\label{fig:image OII}
\end{figure*}

\begin{figure*}
\includegraphics[width=0.60\textwidth]{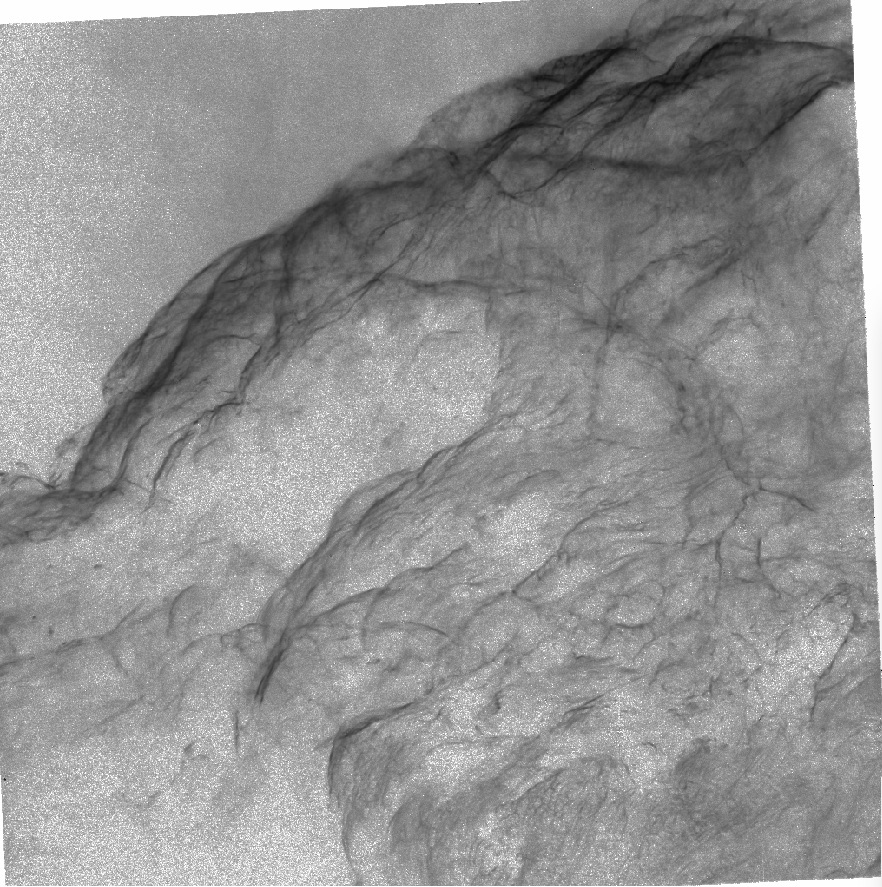}
\caption{Same as Fig. \ref{fig:image OII}, for \OIII $\lambda$5007.}
\label{fig:image OIII}
\end{figure*}

\begin{figure*}
\includegraphics[width=0.60\textwidth]{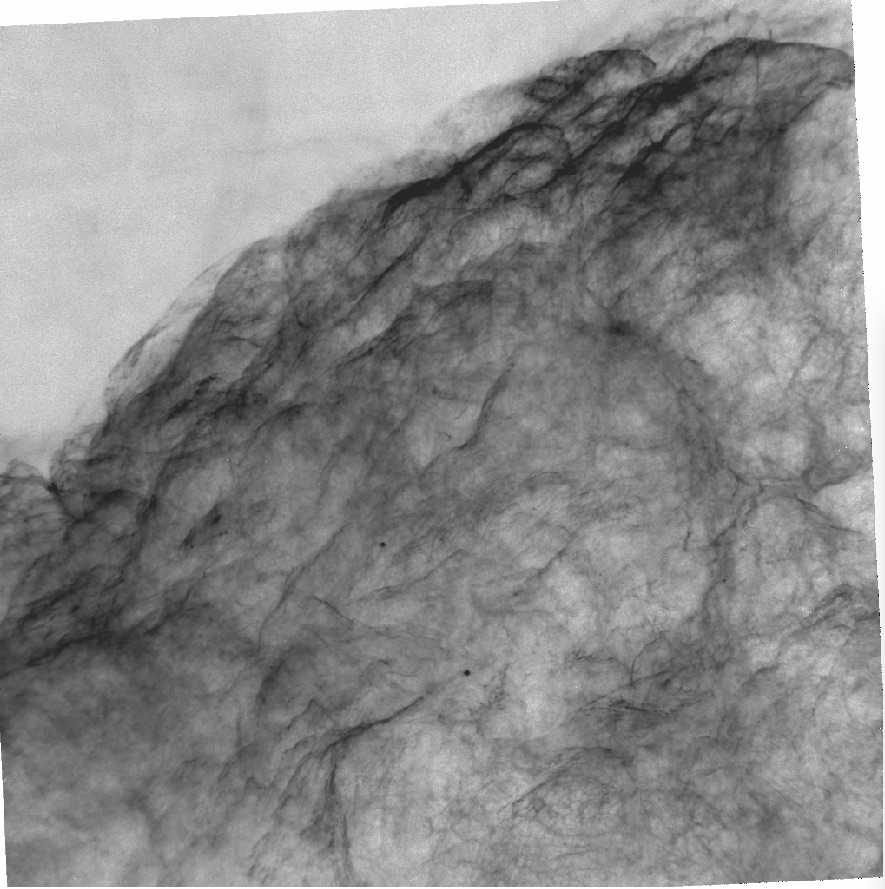}
\caption{Same as Fig. \ref{fig:image OII}, for H$\alpha$. Note the presence of two emission-line stars near the central part of the image.}
\label{fig:image Ha}
\end{figure*}

\begin{figure*}
\includegraphics[width=0.60\textwidth]{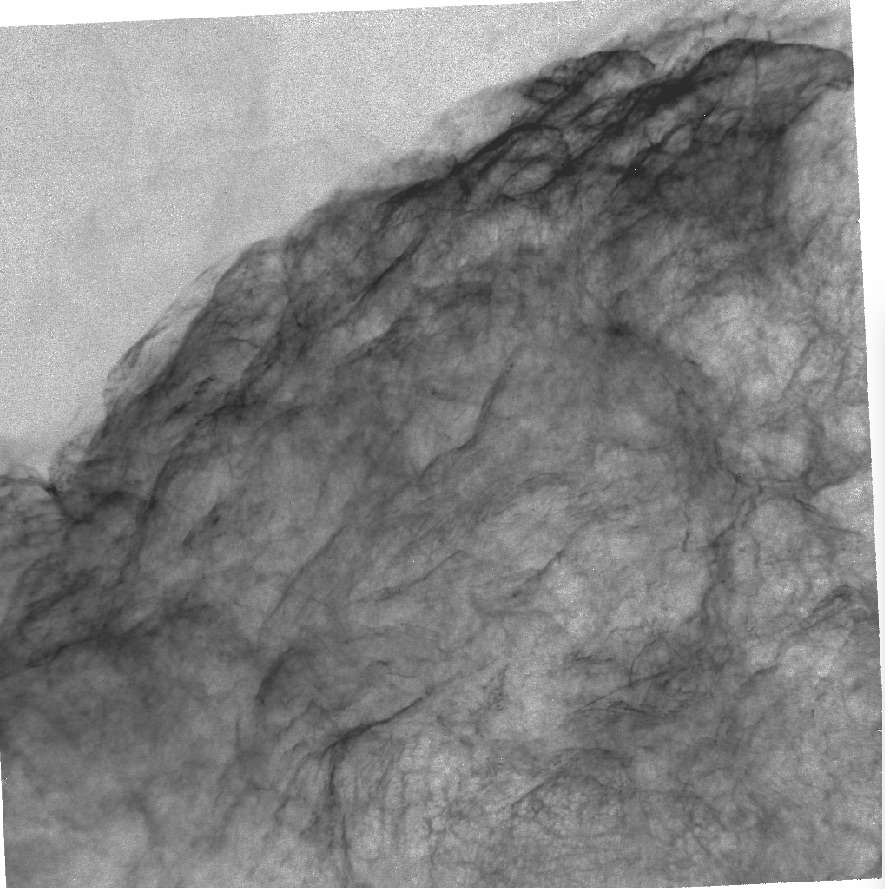}
\caption{Same as Fig. \ref{fig:image OII}, for \NII $\lambda$6583.}
\label{fig:image NII}
\end{figure*}

\begin{figure*}
\includegraphics[width=0.60\textwidth]{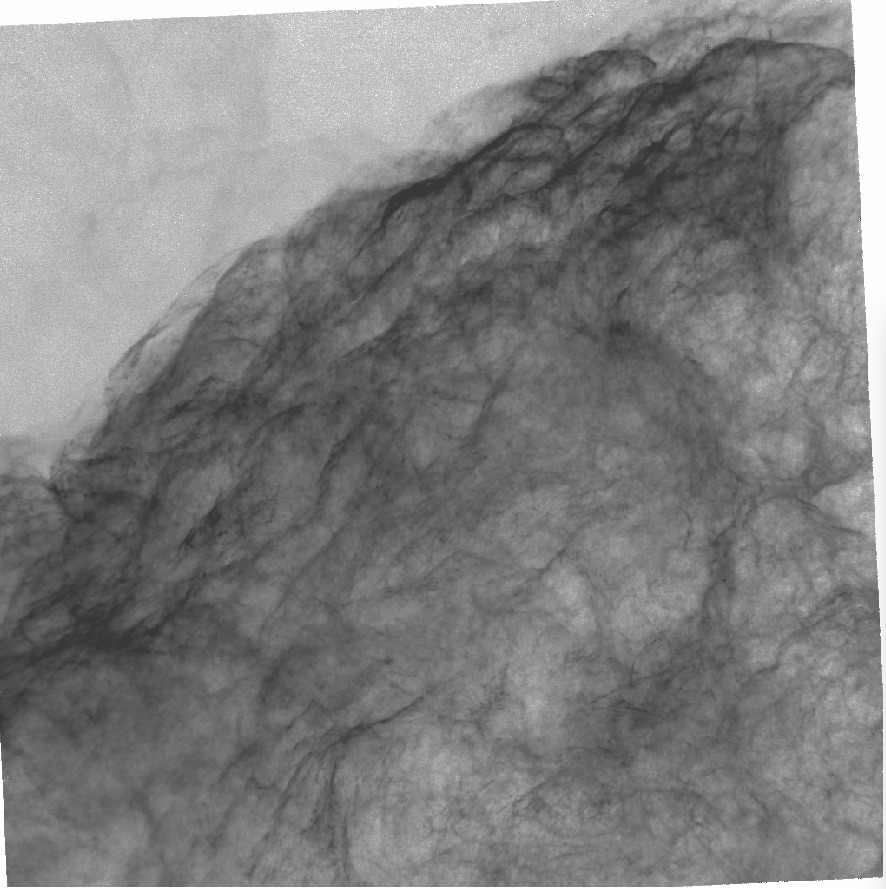}
\caption{Same as Fig. \ref{fig:image OII}, for \SII $\lambda$6716+$\lambda$6731.}
\label{fig:image SII}
\end{figure*}

\begin{figure*}
\includegraphics[width=0.66\textwidth]{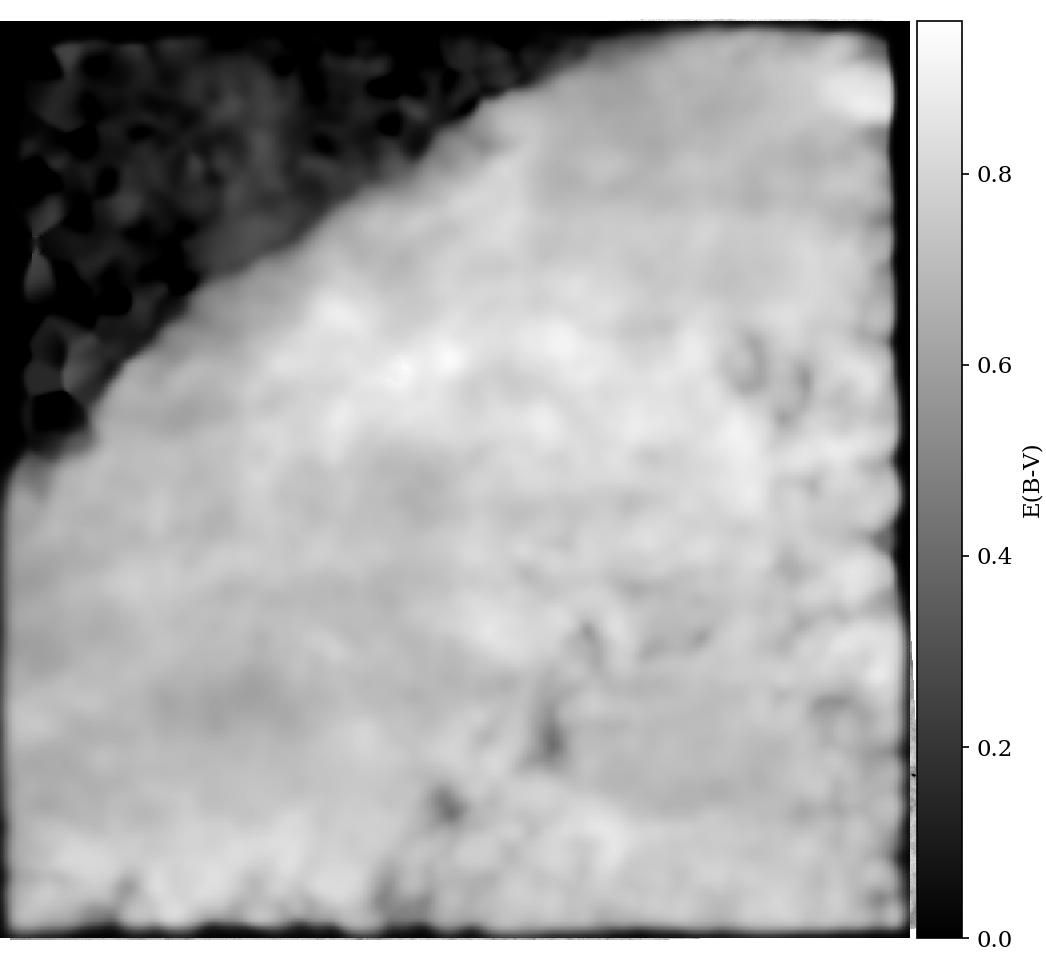}
\caption{Extinction map of north-eastern part of IC 443 using the \Ha/\Hb line ratio. The map was smoothed with a Gaussian distribution with $\sigma$= 5.0 arcsec. }
\label{fig:extinction_map}
\end{figure*}

\begin{figure*}
\centering
\includegraphics[width=0.67\textwidth]{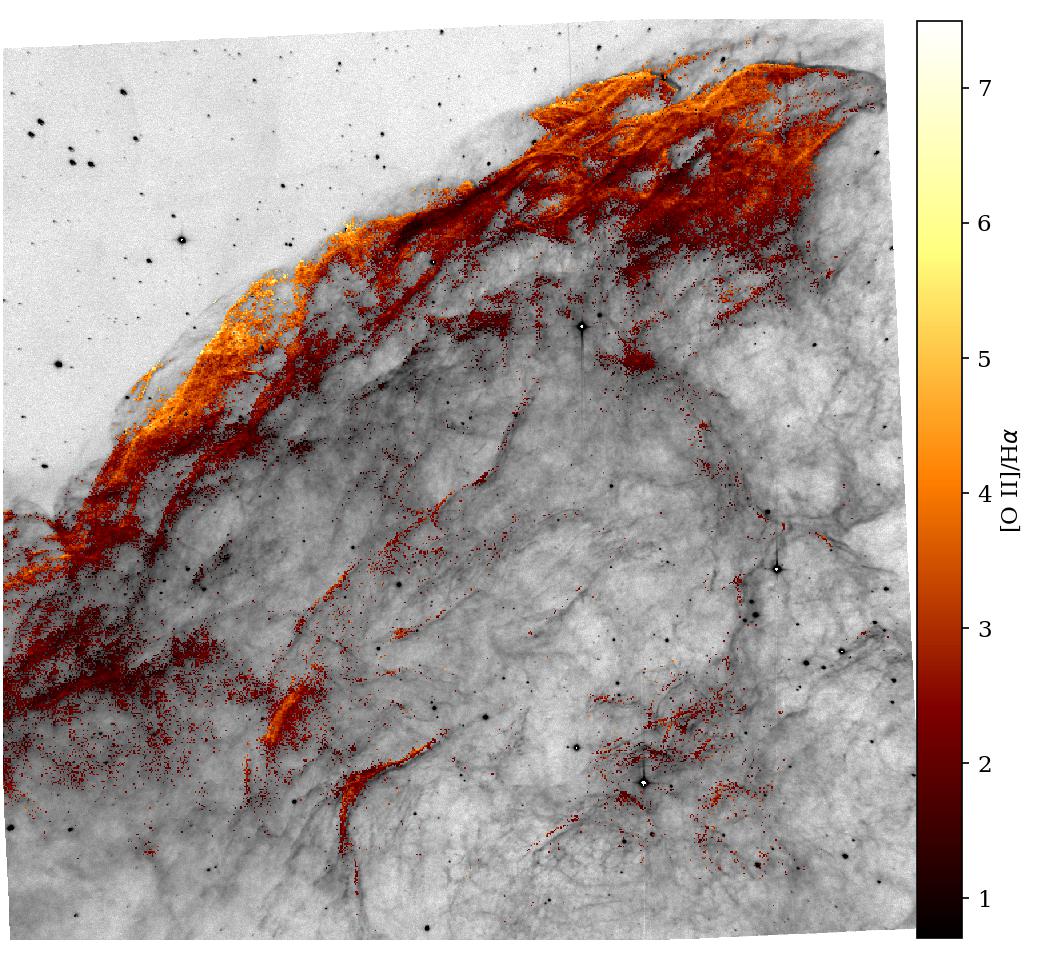}
\caption{Map of the \OII $\lambda$3727+$\lambda$3729/\Ha line ratio. Background image : \Ha emission.}
\label{fig:colormap_OII_Ha}
\end{figure*}

\begin{figure*}
\centering
\includegraphics[width=0.67\textwidth]{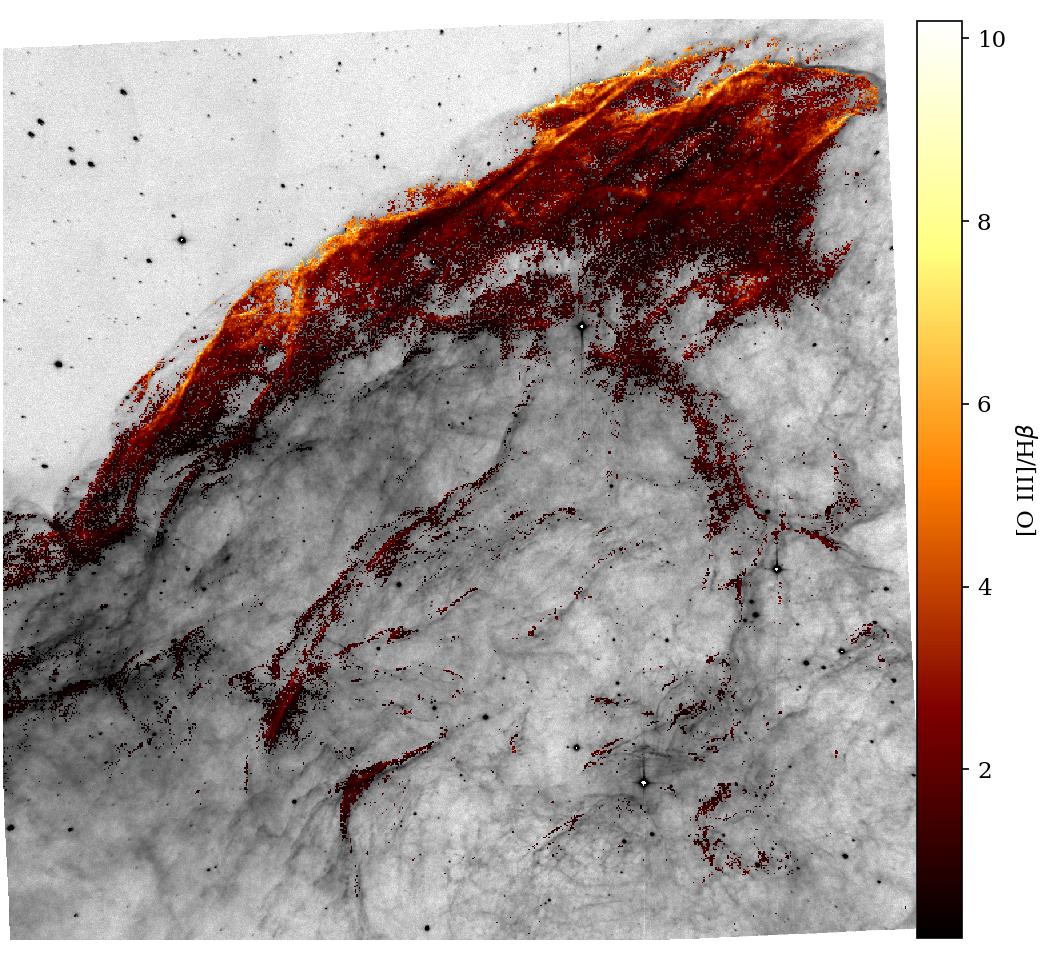}
\caption{Same as Fig. \ref{fig:colormap_OII_Ha}, for \OIII $\lambda$5007/\Hb.}
\label{fig:colormap_OIII_Hb}
\end{figure*}

\begin{figure*}
\centering
\includegraphics[width=0.67\textwidth]{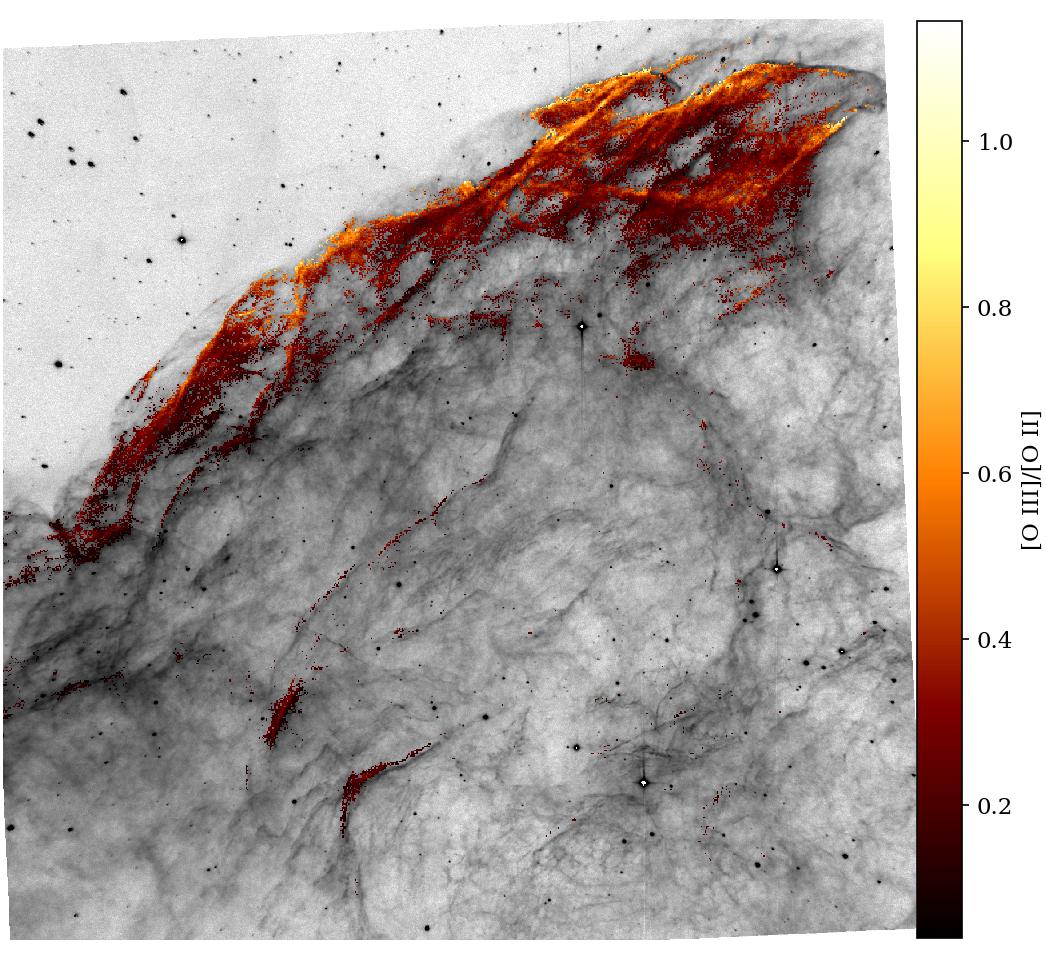}
\caption{Same as Fig. \ref{fig:colormap_OII_Ha}, for \OIII $\lambda$5007/\OII $\lambda$3727+$\lambda$3729.}
\label{fig:colormap_OIII_OII}
\end{figure*}

\begin{figure*}
\centering
\includegraphics[width=0.67\textwidth]{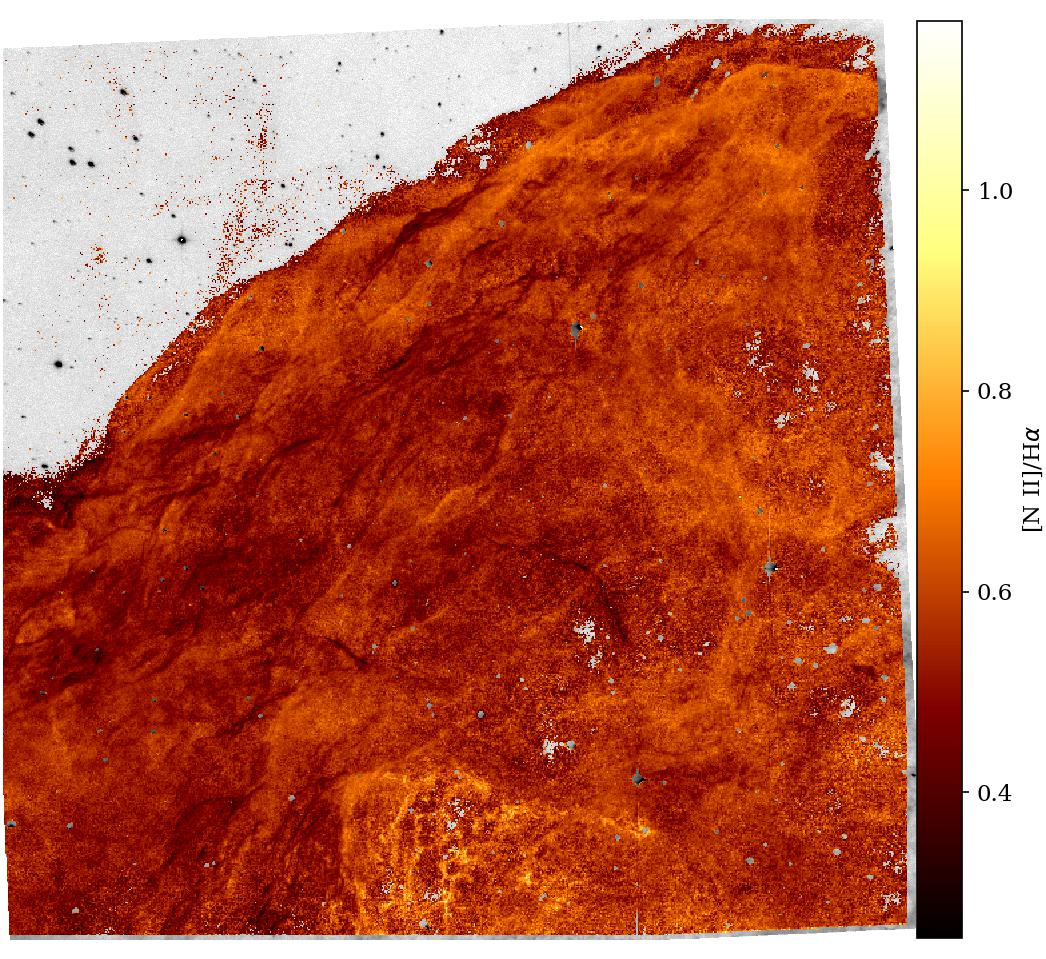}
\caption{Same as Fig. \ref{fig:colormap_OII_Ha}, for \NII $\lambda$6583/\Ha.}
\label{fig:colormap_NII_Ha}
\end{figure*}

\begin{figure*}
\centering
\includegraphics[width=0.67\textwidth]{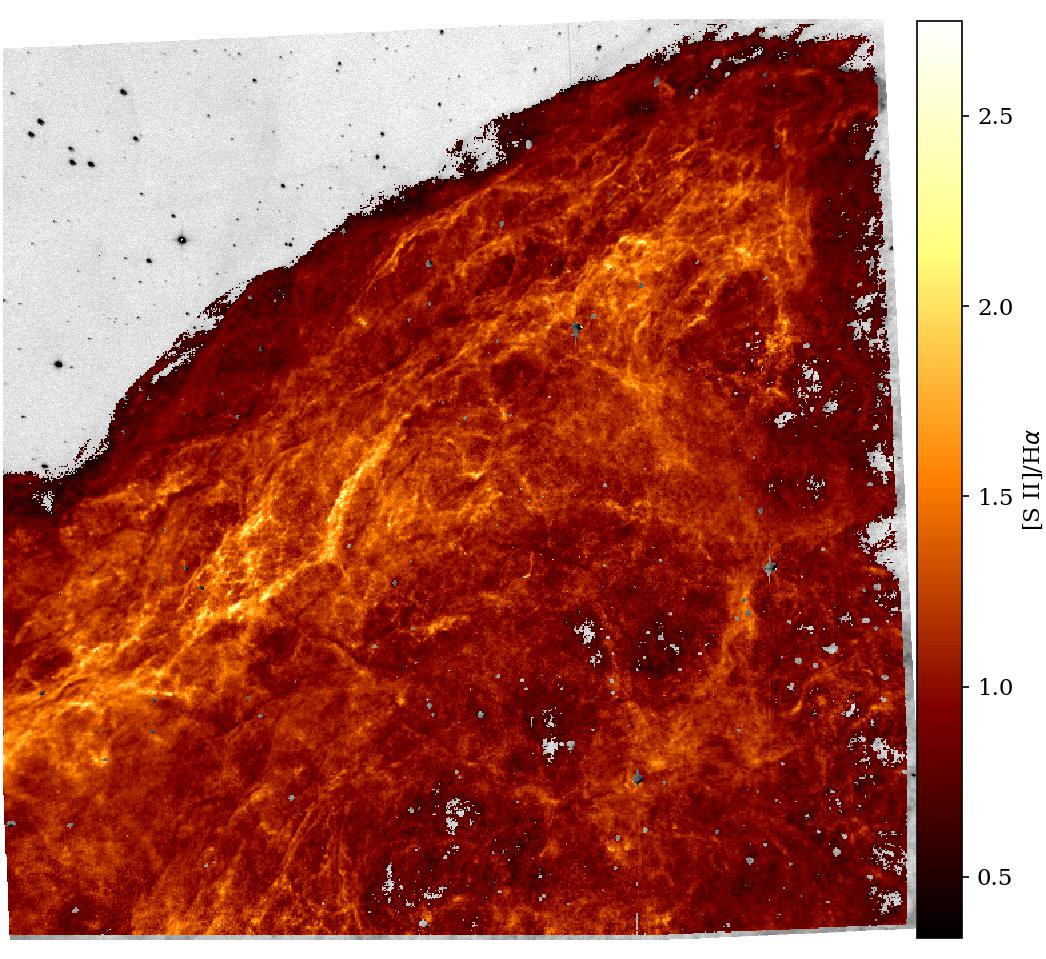}
\caption{Same as Fig. \ref{fig:colormap_OII_Ha}, for \SII $\lambda$6716+$\lambda$6731/\Ha.}
\label{fig:colormap_SII_Ha}
\end{figure*}

\begin{figure*}
\centering
\includegraphics[width=0.67\textwidth]{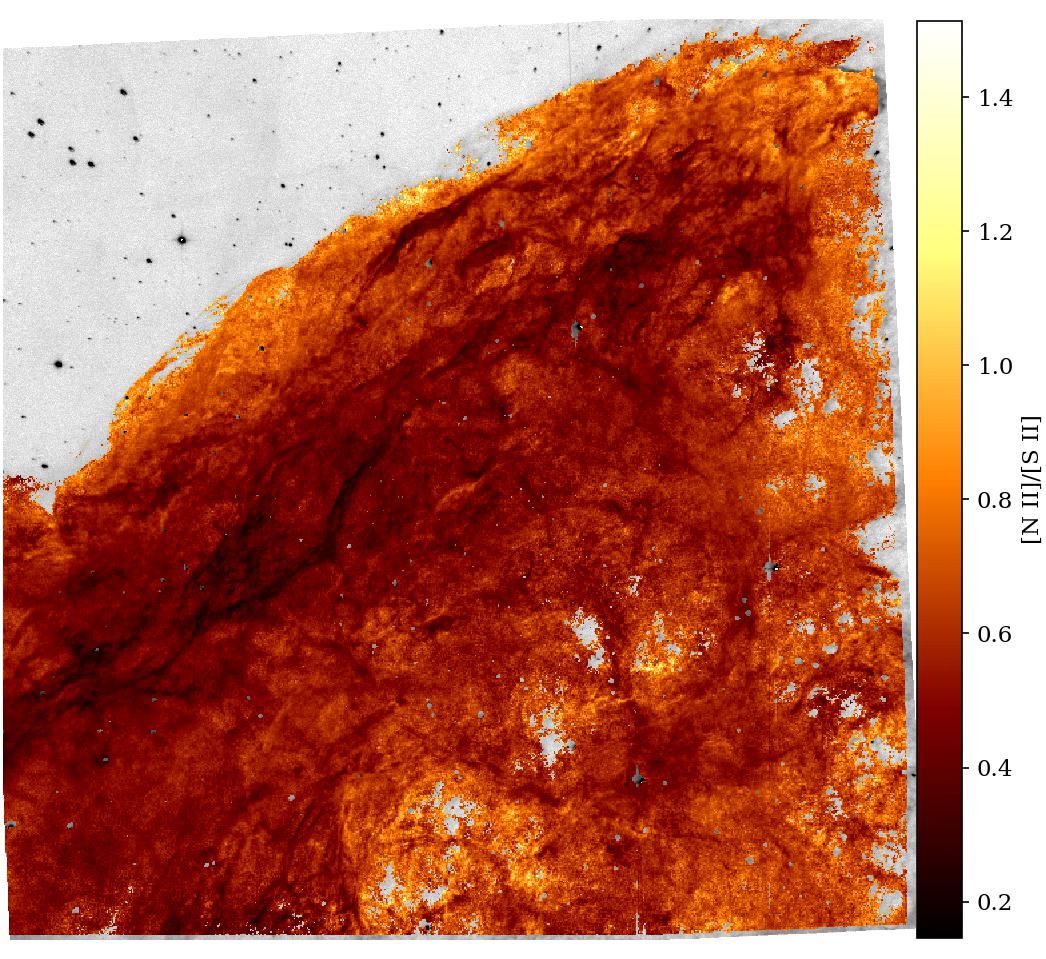}
\caption{Same as Fig. \ref{fig:colormap_OII_Ha}, for \NII $\lambda$6583/\SII $\lambda$6716+$\lambda$6731.}
\label{fig:colormap_NII_SII}
\end{figure*}

\begin{figure*}
\centering
\includegraphics[width=0.67\textwidth]{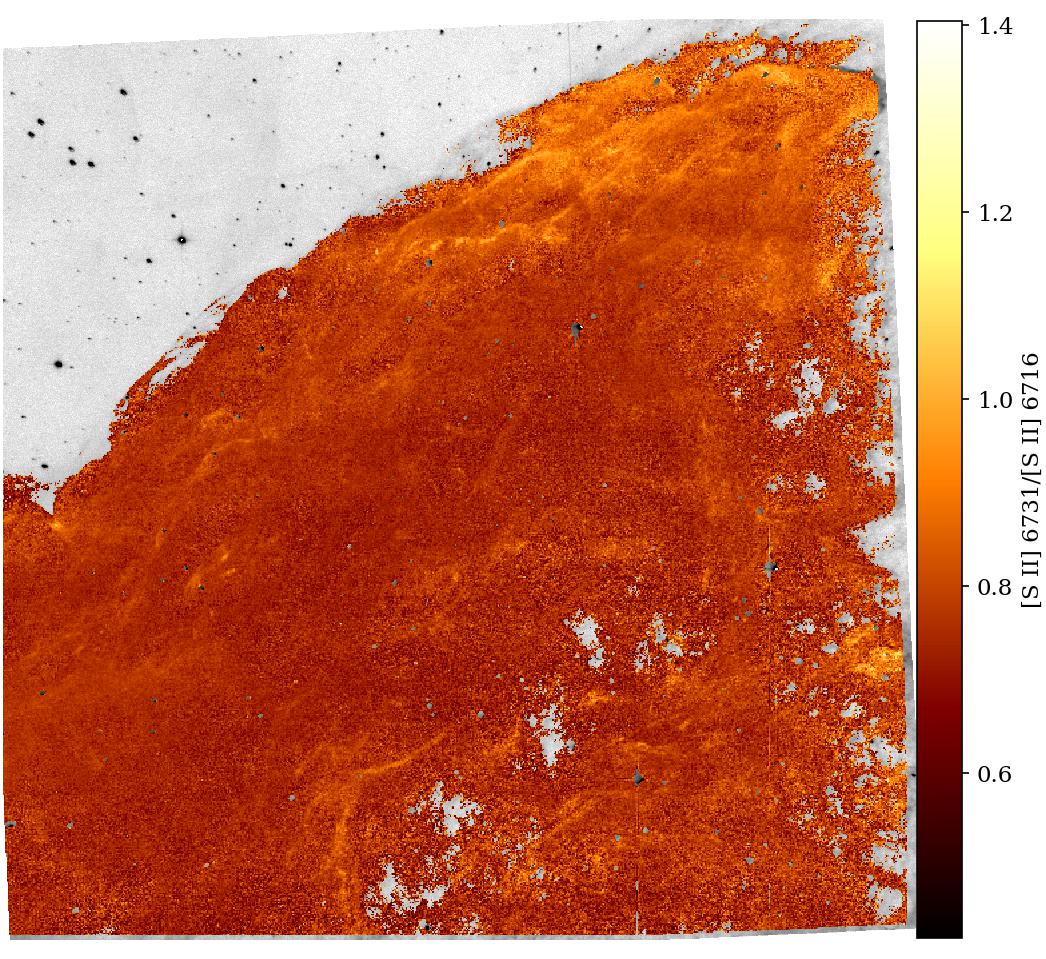}
\caption{Same as Fig. \ref{fig:colormap_OII_Ha}, for \SII $\lambda$6731/\SII $\lambda$6716.}
\label{fig:colormap_SII_SII}
\end{figure*}

\bsp	
\label{lastpage}
\end{document}

%% file: table1.tex
\begin{table*} 
  \caption{Characteristics of the multispectral cubes}
  \label{tab:observations}
  \begin{center}
    \leavevmode
    \begin{tabular}{c c c c c c} 
      \hline 
      \hline              
      Characteristics & SN1 & SN2 & SN3 \\  
      \hline   
      Spectral range (\AA) & 3650-3850 & 4800-5200 & 6510-6850 \\ 
      Spectral resolution (R) & 500 & 800 & 1500  \\
      Exposure time per image (s) & 30 & 30 & 20 \\ 
      Number of images & 88 & 180 & 266 \\ 
      Total observing time & 44 min & 1h30m & 1h28m40s\\ 
      \hline
    \end{tabular}
  \end{center}
\end{table*}

%% file: table2.tex
\begin{table} 
  \caption{Solar abundances with respect to hydrogen used during model calculation based on the compilation of Asplund et al. 2009, Scott et al. 2014 and Grevesse et al. 2014. The abundance of nitrogen and sulfur was modified according to our observations and explained in section 3.2.5.}
  \label{tab:abundances}
  \begin{center}
    \leavevmode
    \begin{tabular}{l l l} 
      \hline 
      \hline              
      Atom \# & Abundances & log(X/H) \\  
      \hline   
      1 & Hydrogen & 0 \\ 
      2 & Helium & -1.07 \\ 
      3 & Lithium & -10.95 \\
      4 & Beryllium & -10.62 \\
      5 & Boron & -9.30 \\
      6 & Carbon & -3.57 \\
      7 & Nitrogen & -3.95 \\ 
      8 & Oxygen & -3.31 \\
      9 & Fluorine & -7.44 \\
      10 & Neon & -4.07 \\
      11 & Sodium & -5.79 \\
      12 & Magnesium & -4.41 \\
      13 & Aluminum & -5.57 \\
      14 & Silicon & -4.49 \\
      15 & Phosphorus & -6.59 \\
      16 & Sulfur & -4.50 \\
      17 & Chlorine & -6.50 \\
      18 & Argon & -5.60 \\
      19 & Potassium & -6.96 \\
      20 & Calcium & -5.68 \\
      21 & Scandium & -8.84 \\
      22 & Titanium & -7.10 \\
      23 & Vanadium & -8.11 \\
      24 & Chromium & -6.38 \\
      25 & Manganese & -6.58 \\
      26 & Iron & -4.53 \\
      27 & Cobalt & -7.07 \\
      28 & Nickel & -5.80 \\
      29 & Copper & -7.82 \\
      30 & Zinc & -7.44 \\
      \hline
    \end{tabular}
  \end{center}
\end{table}

%% file: table3.tex
\begin{table} 
  \caption{Shock parameters used to compute a grid of 57032 models combined with abundances given in table \ref{tab:abundances}.}
  \label{tab:shock_parameters}
  \begin{center}
    \leavevmode
    \begin{tabular}{l l l} 
      \hline 
      \hline              
      Shock parameters & Value range & Interval \\  
      \hline   
      Shock velocity & (20 - 150) km s$^{-1}$ & 1 km s$^{-1}$ \\ 
      Pre-shock density & (14 - 60) cm$^{-3}$ & 2 cm$^{-3}$ \\ 
      Cut-off temperature & (1000 - 20000) K & 500 K \\ 
      Magnetic field & 1 \muG for all model &  \\ 
      \hline
    \end{tabular}
  \end{center}
\end{table}